\documentclass[preprint,showpacs]{revtex4}
\usepackage{amsmath}
\usepackage{amsfonts}
\usepackage{graphicx}
\usepackage{epsfig}
\begin{document} 
\title{On quantum mechanical transport coefficients in nonequilibrium nuclear processes with application to heavy-ion collisions}
\author{Yamen Hamdouni\footnote{Enseignant Vacataire}}
\affiliation{D\'epartement de Pharmacie, Universit\'e Mentouri de Constantine, Constantine, Algeria}
\begin{abstract}
The  elements of the quantum mechanical diffusion matrix, leading to a Gibbs  equilibrium state for a set of $N$ coupled quantum harmonic oscillators are derived within Lindblad's axiomatic approach. Consequences of the fundamental constraints on the quantum friction coefficients are discussed. We derive the equations of motion for the expectation values and variances, and we solve them analytically. We apply our results to the description of the charge and mass asymmetry coordinates in heavy-ion collisions, and we investigate the effect of dissipation on tunneling in sub-barrier processes.  
\end{abstract}

\maketitle
\section{Introduction}
Over the last decades, there has been an increasing interest in  dissipative phenomena in heavy-ion collisions and nuclear fission~\cite{hof1}.  The description of these  processes may be achieved  by the  characterization of  the dynamics of certain collective degrees of freedom such as the relative motion, shape deformation, and  mass and charge asymmetries~\cite{fiss}. The energy dissipation occurs mainly as a result of the coupling of the slow collective degrees of freedom to the fast intrinsic degrees of freedom (e.g. nucleons motion). The latter may be regarded as a heat bath in thermal equilibrium at each stage of the reaction~. 

Classically, the process of dissipation is quite well understood; a typical example is the  Brownian motion which has attracted much attention since Eisentien's  seminal work. The conventional way of treating dissipation and damping of classical systems may be fulfilled by a Langevin equation involving  frictional and   fluctuation forces~\cite{fiss,zwanz}. Another equivalent approach is the Fokker-Planck equation~\cite{fock}. The latter constitutes a probabilistic description  dealing, mainly,  with evolved distribution functions. Here we should like to mention that the above approaches have been successfully applied to the fission process by considering the deformation dynamics of atomic nuclei as  a kind of  Brownian motion~\cite{Wu1}. 

There exist, on the other hand, strong experimental evidences revealing that at low energies, the quantal effects are important at least in the early stage of the reaction~\cite{berlan}, as is the case for the fast charge equilibration in heavy-ion collisions~\cite{fast}. The description of dissipation phenomena at the quantum level is undoubtedly of fundamental significance, in particular at low temperatures where the quantum nature of the physical processes play a crucial role~\cite{weiss}. It turned out that the direct quantization of the classical Langevin equation is quite problematic. Indeed, the usual quantization procedure can be carried out provided the equations of motion  result from the classical Hamilton's principle. For this to be the case for the Langevin equation, the Lagrangian (Hamiltonian) should be explicitly time dependent. A first attempt in this direction consisted in the introduction of a time-dependent mass, which reproduces, in the classical limit, the frictional force~\cite{mass}. It has been quickly recognized that this method exhibits fundamental difficulties.

Later a nonlinear Shr\"{o}dinger equation has been proposed by Kostin in order to describe  dissipation in quantum systems. The above equation reads~\cite{kostin}
\begin{equation}
i\hbar\frac{\partial\Psi}{\partial t}=\Biggl[\hat H+i\gamma\Bigl(\ln\frac{\Psi^*}{\Psi}-\Bigl\langle \ln\frac{\Psi^*}{\Psi}\Bigl\rangle\Bigl)\Biggl]\Psi,\label{kostin}
\end{equation}
where $\hat H$ is the Hamiltonian of the free system (i.e without dissipation),  $\gamma$ is the friction coefficient, and $\langle.\rangle$ denotes the expectation value. The origin of this equation can be elegantly explained using the Madelung hydrodynamical interpretation of  Shr\"{o}dinger equation: The gradient of the phase amplitude, given by the logarithm of the ratio of the wave function to its complex conjugate, is inversely  proportional to the irrotational  velocity field of the  probability fluid. In addition to the  nonlinearity of equation~(\ref{kostin}), which violates the superposition principle, its generalization   to several degrees of freedom is not straightforward.

It turns out that the natural way of describing dissipation phenomena in quantum mechanics consists in dealing with the system as coupled to a large heat bath  such that  irreversible energy flows,  from the former to the latter, take place~\cite{zwanz, weiss, gardiner,petr}.  Hence, the usual unitary group description of the evolution of quantum systems is not suitable  for dealing with such processes  since  the irreversibility  effectively introduces a preferable direction of the  time.  The latter fact  can be accounted for using the notion of dynamical semi-groups, which are the generalization of unitary groups to non-Hamiltonian systems. This is, in particular, the case when the system of interest is opened, that is, when it is coupled to  external systems or  regarded as a part of a larger system. Recall that for a trace-preserving one-parameter semi-group,  the group condition is merely replaced by:  $\Phi_{t+s}=\Phi_t \circ \Phi_s, \quad t>s$~\cite{semigroup};  the evolution equation for the reduced density matrix in the Schr\"odinger picture becomes
\begin{equation}
\frac{d\hat\rho(t)}{dt}=-\frac{i}{\hbar}[\hat H,\hat\rho(t)]+\mathcal D(\hat\rho(t)),\label{mas1}
\end{equation}
where $\hat H$ is the Hamiltonian of the system, and $\mathcal D$ is a superoperator usually called the dissipator. Here we should like to mention that the coupling between the system and the heat reservoir is generally assumed  weak so that the Markovian approximation is  applicable~\cite{weiss}. This approximation is  often used in quantum optics~\cite{gardiner}, and in studying  nuclear fission~\cite{fiss}. 

The most general form of the dissipator $\mathcal D$  giving rise to a quantum mechanical Markovian master equation was found by Lindblad~\cite{lind1,lind2}; it is given by
\begin{equation}
\mathcal D(\hat\rho(t))=\frac{1}{2\hbar}\sum_\ell([\hat V_\ell \hat\rho(t),\hat V_\ell^\dag]+[\hat V_\ell, \hat\rho(t)\hat V_\ell^\dag]), \label{mas2}   
\end{equation}        
where the $\hat V_\ell$ are called Lindblad's operators; they depend on the system's variables and they model the effect of the environment. The dual  evolution operator to~(\ref{mas1}) describes the  time development  of any Heisenberg operator $\hat A$, namely, 
\begin{equation}
 \frac{d\hat A(t)}{dt}=\frac{i}{\hbar}[\hat H,\hat A(t)]+\frac{1}{2\hbar}\sum_\ell([\hat V_\ell^\dag[ \hat A(t),\hat V_\ell]+[\hat V_\ell^\dag, \hat A(t)]\hat V_\ell]).\label{mas3}
 \end{equation}

Depending on the system of interest, one can obtain axiomatically  the master equation  describing its damping by  properly  choosing the form of the Lindblad's operators. This axiomatic approach has been extensively applied to the one-dimensional harmonic oscillator because of its simplicity~\cite{sandu,isar1,isar2,antonenko,isar3,isar4,isar5,adam1,adam2,genkin1}. The resulting master equation involves diffusion coefficients in  coordinate and momentum which are of quantum nature~\cite{adam3} because of the fundamental constraints they have to satisfy. The-two dimensional case has been considered in~\cite{two1}, and~\cite{two2}. It is, however, generally assumed that the off-diagonal elements of the quantum mechanical diffusion matrix are negligible. This, certainly, cannot be the case  when the harmonic oscillators are strongly interacting with each other. In this paper we  apply Lindblad's axiomatic approach  to  explicitly derive  the elements of the diffusion matrix, for the general case of $N$ harmonic oscillators, by imposing the condition that the asymptotic state  is a Gibbs state. We should like to stress that the  system we are considering here is of significant practical importance in nuclear physics, since the nuclear potential can be approximated, at least locally, by harmonic oscillators. In section~\ref{sec2} we present a detailed derivation of the elements of the diffusion matrix and we investigate the constraints that should be satisfied  by the transport coefficients. Section~\ref{sec3} is devoted to the equations of motion for the expectation values and variances of the coordinates an momenta operators. In Section~\ref{sec4} we apply our results to the description of fast charge equilibration in deep-inelastic collisions, and we investigate the effect of dissipation on tunneling in sub-barrier processes. We end the paper with a summary.

\section{Quantum mechanical transport coefficients for $N$ coupled harmonic oscillators\label{sec2}}
The aim of this section is the derivation of the quantum mechanical diffusion coefficients in (mixed) coordinates and momenta for $N$ coupled quantum harmonic oscillators by imposing a Gibbs asymptotic state for the evolved reduced density matrix. 

\subsection{Explicit derivation}
The most general form of a $N$-dimensional quadratic Hamiltonian in coordinates and momenta is given by
\begin{eqnarray}
\hat H=\sum\limits_{k=1}^N \Bigl( \frac{\hat{p}_k^2}{2M_k}+ \frac{1}{2}M_k\Omega_k^2 \hat{q}_k^2&+&\frac{\mu_{kk}}{2}(\hat{ p}_k\hat{q}_k+  \hat{q}_k\hat{p}_k)\nonumber\Bigl) \\ &+& \frac{1}{2}\sum\limits_{k\neq j}^N(\nu_{kj} \hat{q}_k\hat{q}_j+\kappa_{kj}\hat{p}_k\hat{p}_j)+ \sum\limits_{k\neq j}^N \mu_{kj} \hat{p}_k \hat{q}_j,
\end{eqnarray}
where the operators $\hat{q}_k$ and $\hat{p}_j$ satisfy the usual canonical commutation relations:
\begin{equation}\label{can}
[\hat q_k,\hat p_j]=i\hbar\delta_{kj},\qquad [\hat q_k,\hat q_j]=[\hat p_k,\hat p_j]=0.
\end{equation}
To ensure that the above Hamiltonian is physical the coupling strengths $\mu_{kj}$, $\nu_{kj}$ and $\kappa_{kj}$ should satisfy certain conditions, some of which will be discussed later in  particular cases. Here we just mention that,   according to Onsager principle~\cite{onsager}, the last two  coefficients should be symmetrical, that is:
\begin{equation}
\nu_{kj}=\nu_{jk}, \qquad \kappa_{kj}=\kappa_{jk}.
\end{equation}

We require  that the system relaxes to a steady  state corresponding to  $N$ independent quantum harmonic oscillators in thermal equilibrium at temperature $k_B T=1/\beta$. This corresponds to complete thermalization of the system. Explicitly we have
 \begin{equation}
\hat \rho_{\rm{eq}}=\exp(-\beta \hat H_{\rm{eq}})/Z,\label{gibbs}
 \end{equation}
 where 
 \begin{equation}
 \hat H_{\rm{eq}}=\sum\limits_{k=1}^N \Bigl( \frac{\hat{p}_k^2}{2m_k}+ \frac{1}{2}m_k\omega_k^2 \hat{q}_k^2\Bigr).
 \end{equation}
 For the sake of generality, we assume that $m_k$ and $\omega_k$ are different from, respectively, $M_k$ and $\Omega_k$. The partition function, $Z$,  can be easily calculated by rewriting $\hat H_{\rm{eq}}$ in terms of the creation and annihilation operators, and then by taking the trace in the occupation  number space; one finds that   
 \begin{equation}
 Z=\prod\limits_{k=1}^N \frac{1}{1-e^{-\hbar \beta \omega_k}}.
 \end{equation} 
For convenience, we further introduce the following operators: 
\begin{align}
\hat T_k^1&=\frac{\hat{p}_k^2}{2m_k}- \frac{1}{2}m_k\omega_k^2 \hat{q}_k^2,\\
\hat T_k^2&= \frac{\omega_k}{2}( \hat{q}_k \hat{p}_k+\hat{p}_k \hat{q}_k),\\
\hat T_k^3&=\frac{\hat{p}_k^2}{2m_k}+ \frac{1}{2}m_k\omega_k^2 \hat{q}_k^2.
\end{align}
Using the canonical commutation relations~(\ref{can}), we can show that the above operators are the generators of an  $SO(2,1)$ Lie group, namely,
\begin{align}
[\hat T^1_k,\hat T^2_j]=-2 i\hbar \omega_k \hat T_k^3 \delta_{kj}, & \qquad  [\hat T^2_k,\hat T^3_j]=2 i\hbar \omega_k \hat T_k^1 \delta_{kj} , \nonumber \\   [\hat T^3_k, \hat T^1_j]= & 2 i\hbar \omega_k \hat T_k^2 \delta_{kj}. \label{lie}
\end{align}
Now, the  Hamiltionian $\hat H$  and the density matrix $\hat\rho_{\rm{eq}}$ may be expressed in terms of the operators $T_k^\ell$ as
\begin{eqnarray}
\hat H&=&\sum\limits_{k=1}^N \Bigl(\epsilon_k \hat T_k^3+ \delta_k \hat T_k^1+ f_k \hat T^2_k\Bigr)+\frac{1}{2}\sum\limits_{k\neq j}^N(\nu_{kj} \hat{q}_k\hat{q}_j+\kappa_{kj}\hat{p}_k\hat{p}_j)+ \sum\limits_{k\neq j}^N \mu_{kj} \hat{p}_k \hat{q}_j\\
\hat H_{\rm{eq}}&=& \sum\limits_{k=1}^N \hat T^k_3,
\end{eqnarray}
with
\begin{eqnarray}
\epsilon_k=\frac{1}{2}\Bigl(\frac{m_k}{M_k}+\frac{M_k\Omega_k^2}{m_k\omega_k^2}\Bigl),&& \quad  \delta_k=\frac{1}{2}\Bigl(\frac{m_k}{M_k}-\frac{M_k\Omega_k^2}{m_k\omega_k^2}\Bigl),\nonumber \\
 f_k&=&\frac{\mu_{kk}}{\omega_k}.
\end{eqnarray}

Since $\hat\rho_{\rm{eq}}=\exp(-\beta \sum_k \hat T^3_k)/Z$ is a  solution of the master equation~(\ref{mas1}) we should observe the following equality:
\begin{equation}
\mathcal{L}[\hat H_S]-\hat H_S+i\sum_\ell\Bigl(\mathcal{L}[\hat V_\ell]\hat V_\ell^\dag-\frac{1}{2}\mathcal{L}[\hat V_\ell^\dag \hat V_\ell]-\frac{1}{2}\hat V_\ell^\dag \hat V_\ell \Bigr)=0\label{two},
\end{equation}
where we have introduced the superoperator $\mathcal{L}$ whose action is defined by
\begin{eqnarray}
\mathcal{L}[\hat A]&=&\exp\Bigl(\beta \sum_k \hat T^3_k\Bigr) \hat A \exp\Bigl(-\beta \sum_k \hat T^3_k\Bigr)\nonumber\\
&=&\hat A+\sum_k\Bigl\{-\frac{\beta}{1!}[\hat A,\hat T_k^3]+\frac{\beta^2}{2!}[[\hat A,\hat T_k^3],\hat T_k^3]-\frac{\beta^3}{3!}[[[\hat A,\hat T_k^3],\hat T^3_k],\hat T_k^3]+\cdots\Bigr\}.\label{super}
\end{eqnarray}
Taking into account the commutation relations~(\ref{lie}), a straightforward application of the above equation yields: 
\begin{eqnarray}
\mathcal{L}[\hat T^1_j]=\cosh(2 \hbar \beta \omega_j)\hat T^1_j+i\sinh(2 \hbar \beta \omega_j)\hat T^2_j\label{com1},\\
\mathcal{L}[\hat T^2_j]=\cosh(2 \hbar \beta \omega_j)\hat T^2_j-i\sinh(2 \hbar \beta \omega_j)\hat T^1_j\label{com2}.
\end{eqnarray}

The Lindblads's operators $\hat V_\ell$  may be expressed, in our case, as linear combinations of the coordinates and momenta operators. This is, in some sense, the quantum analogue of Hooke's law in classical mechanics. Consequently, we can write:
\begin{equation}
\hat V_\ell=\sum_j^N (a^\ell_j \hat q_j+b_j^\ell \hat p_j),\quad \hat V^\dag_\ell=\sum_j^N (a^{\ell*}_j \hat q_j+b_j^{\ell*} \hat p_j), \qquad \ell=\overline{1,2N},  
\end{equation}
where $ a^\ell_j$ and $b_j^\ell$ are complex numbers.

By making use of the canonical commutation relations (\ref{can}), together with equations~(\ref{super}) and~(\ref{com1})-(\ref{com2}), it can be shown that  
\begin{eqnarray}
\mathcal{L}[\hat H_S]-\hat H_S&=&\sum_k\Bigl\{\bigl[\delta_k [\cosh(2 \hbar \beta \omega_k)-1]-i f_k\sinh(2 \hbar \beta \omega_k \bigr] \hat T^1_k\nonumber \\&+&\bigl[f_k[\cosh(2 \hbar \beta \omega_k)-1]+i \delta_k \sinh(2 \hbar \beta \omega_k)\bigr]\hat T^2_k\Bigr\}\nonumber\\&+&\frac{1}{2}\sum_{k\neq j}\Bigl[\nu_{kj}(\cosh(\hbar \beta \omega_k) \cosh(\hbar \beta \omega_j)-1)-\kappa_{kj}m_k m_j\omega_k\omega_j \nonumber\\&\times& \sinh(\hbar \beta \omega_k) \sinh(\hbar \beta \omega_j)+i\mu_{kj} m_k\omega_k \sinh(\hbar \beta \omega_k)\cosh(\hbar \beta \omega_j)\Bigl]\hat q_k\hat q_j \nonumber\\
&+&\frac{1}{2}\sum_{k\neq j}\Bigl[\kappa_{kj}(\cosh(\hbar \beta \omega_k) \cosh(\hbar \beta \omega_j)-1)-\frac{\nu_{kj}}{m_k m_j\omega_k\omega_j} \nonumber\\&\times & \sinh(\hbar \beta \omega_k) \sinh(\hbar \beta \omega_j)+i\frac{\mu_{kj}}{ m_k\omega_k} \cosh(\hbar \beta \omega_k)\sinh(\hbar \beta \omega_j)\Bigl]\hat p_k \hat p_j\nonumber\\
&+&\sum_{k\neq j}\Bigl[\mu_{kj}(\cosh(\hbar \beta \omega_k) \cosh(\hbar \beta \omega_j)-1)-\frac{\mu_{jk} m_j \omega_j}{m_k \omega_k} \nonumber\\&\times & \sinh(\hbar \beta \omega_k) \sinh(\hbar \beta \omega_j)+2 i\Bigl(\frac{\nu_{kj}}{m_k\omega_k} \sinh(\hbar \beta \omega_k) \cosh(\hbar \beta \omega_j)\nonumber\\&+&\kappa_{kj}m_j\omega_j \cosh(\hbar \beta \omega_k) \sinh(\hbar \beta \omega_j)\Bigl)\Bigl]\hat p_k\hat q_j
\end{eqnarray}
The other terms  of equation~(\ref{two}) are easily calculated (see the appendix). In fact, the latter equation implies that the coefficients of all the involved operators should be equal to zero. As a result, by introducing the following notations: 
\begin{eqnarray}
D_{q_kq_j}&=&\frac{\hbar}{2}{\rm Re}\sum_\ell a^{\ell*}_k a^\ell_j, \quad  D_{p_kp_j}=\frac{\hbar}{2}{\rm Re}\sum_\ell b^{\ell*}_k b^\ell_j, \label{coef1}\\
D_{q_kp_j}&=&-\frac{\hbar}{2}{\rm Re}\sum_\ell a^{\ell*}_k b^\ell_j, \quad \lambda_{kj}=-{\rm Im}\sum_\ell a^{\ell*}_k b^\ell_j,\label{coef2}\\
\alpha_{kj}&=&-{\rm Im}\sum_\ell a^{\ell*}_k a^\ell_j,\quad \eta_{kj}=-{\rm Im}\sum_\ell b^{\ell*}_k b^\ell_j,\label{coef3}
\end{eqnarray}
 we obtain  three independent sets  of linear algebraic equations. The first one is as follows:
\begin{eqnarray}
 \frac{\hbar\mu_{kk}}{2} [\cosh(2\hbar\beta \omega_{k}) - 1] &=& 
- \Bigl(D_{q_kq_k} m_k \omega_k - \frac{ D_{p_kp_k}}{ m_k \omega_k}\Bigr) 
   \sinh(\hbar\beta \omega_{k}) \cosh(\hbar\beta \omega_{k})\nonumber\\& -& \frac{ D_{p_kp_k}}{ m_k \omega_k} \sinh(\hbar\beta \omega_{k})+ \hbar\lambda_{kk} \Bigl(\cosh(\hbar\beta \omega_{k})+1\Bigr),\\
   \frac{\hbar \mu_{kk}}{2 m_k \omega_k}  \sinh(2\hbar\beta \omega_{k}) &=& - D_{q_kq_k} \Bigl(\cosh(\hbar\beta \omega_{k}) - 1\Bigr)^2 - 
    \frac{ \hbar \lambda_{kk}}{m_k \omega_k} \sinh(\hbar\beta \omega_{k})\nonumber\\ &+& 
    \frac{D_{p_kp_k}}{ m_k^2 \omega_k^2} \sinh^2(\hbar\beta \omega_{k}),\\
   2\hbar \mu_{kk}  m_k \omega_k \sinh(2\hbar\beta \omega_{k}) &=&  D_{p_kp_k} \Bigl(\cosh(\hbar\beta \omega_{k}) - 1\Bigr)^2 +
   \hbar \lambda_{kk} m_k \omega_k \sinh(\hbar\beta \omega_{k})\nonumber\\&-& D_{q_kq_k} m_k^2 \omega_k^2 \sinh^2(\hbar\beta \omega_{k}),\\
   \delta_k\Bigl(\cosh(2\hbar\beta \omega_{k})-1\Bigr)&=&\frac{D_{p_kq_k}}{\hbar\omega_k}\Bigl(1-\cosh(\hbar\beta \omega_{k})\Bigr)\sinh(\hbar\beta \omega_{k}),
\end{eqnarray}  
The second one reads:
\begin{eqnarray}
 &&  m_j \omega_j \Bigl[2 m_k D_{q_kq_j} \omega_k \Bigl(-1 + 
       \cosh(\hbar \beta \omega_k)\Bigr) \Bigl(-1 + \cosh(\hbar \beta \omega_j)\Bigr) \nonumber\\&& + \Bigl( \hbar \lambda_{jk} + \hbar \mu_{jk} \cosh(\hbar \beta \omega_j)\Bigr) \sinh(\hbar \beta \omega_k)\Bigr] +\Bigl( \hbar \lambda_{kj} m_k \omega_k + \hbar \mu_{kj} m_k \omega_k \cosh(\hbar \beta \omega_k)\nonumber\\ &&- 
    2 D_{p_kp_j} \sinh(\hbar \beta \omega_k)\Bigr) \sinh(\hbar \beta \omega_j)=0,\\\nonumber \\
    && 8 D_{p_kp_j}\sinh^2\Bigl(\dfrac{\hbar \beta \omega_k}{2}\Bigr) \sinh^2\Bigl(\dfrac{\hbar \beta \omega_j}{2}\Bigr) + 
  m_j \hbar \omega_j \Bigl(  \lambda_{jk}- 
     \mu_{jk} \cosh(\hbar \beta \omega_k)\Bigr) \sinh(\hbar \beta \omega_j)\nonumber\\ && + 
  m_k \hbar \omega_k \sinh(\hbar \beta \omega_k) \Bigl( \lambda_{kj} - 
     \mu_{kj} \cosh(\hbar \beta \omega_j) - 
     2 m_j D_{q_kq_j} \omega_j \sinh(\hbar \beta \omega_j)\Bigr) = 0,\\ \nonumber \\
     &&\frac{2 D_{p_kp_j}}{m_k\omega_k} \Bigl (\cosh(\hbar \beta \omega_j)-1\Bigr) \sinh(\hbar \beta \omega_k)- 
 2 m_j  \omega_j D_{q_kq_j}\Bigl( 
     \cosh(\hbar \beta \omega_k)-1\Bigr) \sinh(\hbar \beta \omega_j)\nonumber\\&& + 
  \hbar \Bigl\{\mu_{kj} -  \lambda_{kj} \cosh(\hbar \beta \omega_j) + 
     \cosh(\hbar \beta \omega_k) [ \lambda_{kj} - \mu_{kj} \cosh(\hbar \beta \omega_j)] \nonumber\\&&- \frac{
     \mu_{jk} m_j \omega_j \sinh(\hbar \beta \omega_j) \sinh(\hbar \beta \omega_k)}{m_k \omega_k}\Bigr\}=0,
\end{eqnarray}
 Finally, the equations of the third set can be written as:
\begin{eqnarray}
&&-2D_{q_kp_j} \Bigl( \cosh(\hbar \beta \omega_k)-1\Bigr) \Bigl( \cosh(\hbar \beta \omega_j)-1\Bigr) + \hbar\Bigl(\frac{-\eta_{kj} +
      \nu_{kj} \cosh(\hbar \beta \omega_j)}{m_k \omega_k}\Bigr)\nonumber\\&&\times \sinh(\hbar \beta \omega_k) - 
  m_j\hbar \omega_j\Bigl (\alpha_{kj} + 
     \kappa_{kj} \cosh(\hbar \beta \omega_k)\Bigr) \sinh(\hbar \beta \omega_j) -2 \Bigl(\frac{m_j\omega_j}{m_k\omega_k}\Bigr)\nonumber\\&& \times
   D_{q_jp_k} \sinh(\hbar \beta \omega_k) \sinh(\hbar \beta \omega_j) = 0,  \\ \nonumber \\    
   && -2 D_{q_jp_k} \Bigl( \cosh(\hbar \beta \omega_k)-1\Bigr) \Bigl(\cosh(\hbar \beta \omega_j)-1\Bigl) - 
    m_k \hbar \omega_k \Bigl(-\alpha_{kj} + 
       \kappa_{kj} \cosh(\hbar \beta \omega_j)\Bigr)\nonumber\\&& \times \sinh(\hbar \beta \omega_k) +\Bigl( \frac{1}{m_j\omega_j}\Bigr)\Bigl[\hbar\eta_{kj} + 
    \hbar \nu_{kj} \cosh(\hbar \beta \omega_k) -2 
    m_k  \omega_k D_{q_kp_j} \sinh(\hbar \beta \omega_k)\Bigr]\nonumber\\ &&\times \sinh(\hbar \beta \omega_j)=0,\\ \nonumber \\
   && \hbar\nu_{kj} - \hbar\eta_{kj} \cosh(\hbar \beta \omega_j) + 
  m_j \omega_j \Bigl(-2D_{q_jp_k} + 
     \hbar \kappa_{kj} m_k \omega_k \sinh(\hbar \beta \omega_k)\Bigr) \sinh(\hbar \beta \omega_j)\nonumber\\&& - 
  \cosh(\hbar \beta \omega_k)  \Bigl[-\hbar\eta_{kj} +\hbar \nu_{kj} \cosh(\hbar \beta \omega_j) - 
     2 D_{q_j p_k} m_j \omega_j \sinh(\hbar \beta \omega_j\Bigr]\nonumber \\ &&+ 
 2 D_{q_kp_j}m_k  \omega_k \Bigl( 
     \cosh(\hbar \beta \omega_j)-1\Bigr) \sinh(\hbar \beta \omega_k)=0
\end{eqnarray}
From equations~(\ref{coef1}) and~(\ref{coef2}), it is clear that the quantities $D_{q_kq_j}$, $D_{p_kp_j}$, $D_{q_kp_j}$ and $\lambda_{kj}$ are the multidimensional extension of the quantum mechanical diffusion and friction coefficients corresponding to the one-dimensional harmonic oscillator~\cite{sandu}. Solving the above sets of algebraic equations for the diffusion coefficients yields: 
\begin{eqnarray}
D_{q_kq_k}&=&\frac{\hbar}{2}\Biggl(\frac{\lambda_{kk}-\mu_{kk}}{m_k\omega_k}\Biggl)\coth \frac{\hbar \beta \omega_k}{2},\label{co1}\\
D_{p_kp_k}&=&\frac{\hbar}{2} m_k\omega_k(\lambda_{kk}+\mu_{kk})\coth\frac{\hbar \beta \omega_k}{2},\label{co2}\\
D_{p_kq_k}&=&D_{q_kp_k}=-\frac{\hbar}{2} \omega_k\delta_k \coth\frac{\hbar \beta \omega_k}{2}=\frac{\hbar}{4}\Bigl(\frac{M_k\Omega_k^2}{m_k\omega_k}-\frac{m_k \omega_k}{M_k}\Bigl)\coth\frac{\hbar \beta \omega_k}{2},\label{co3}\\
D_{q_kq_j}&=&D_{q_jq_k}=\frac{\hbar}{4}\Biggl(\frac{\lambda_{jk}-\mu_{jk}}{m_k\omega_k}\coth\frac{\hbar \beta \omega_k}{2}+\frac{\lambda_{kj}-\mu_{kj}}{m_j\omega_j}\coth\frac{\hbar \beta \omega_j}{2}\Biggl),\label{co4}\\
D_{p_kp_j}&=&D_{p_jp_k}=\frac{\hbar}{4}\Biggl((\lambda_{jk}+\mu_{jk})m_k\omega_k\coth\frac{\hbar \beta \omega_k}{2}+(\lambda_{kj}+\mu_{kj})m_j\omega_j\coth\frac{\hbar \beta \omega_j}{2}\Biggl),\label{co5}\\
D_{q_kp_j}&=&D_{p_jq_k}=\frac{\hbar}{4}\Biggl(\frac{\eta_{kj}+\nu_{kj}}{m_k\omega_k}\coth\frac{\hbar \beta \omega_k}{2}+(\alpha_{kj}-\kappa_{kj})m_j\omega_j\coth\frac{\hbar \beta \omega_j}{2}\Biggl),\label{co6}
\end{eqnarray}
which constitute the main result of this work. We can clearly see that the   diffusion coefficients~(\ref{co4})-(\ref{co6})  can by no means be neglected when the coupling constants $\nu_{kj}$ and $\kappa_{kj}$ are comparable with, respectively, $m_k\omega_k^2$ and $1/m_k$ (i.e. strong coupling). We shall discuss later the influence of these constants on the evolution in time of the relevant physical quantities. 

 The  diffusion coefficients form a $N\times N$ symmetrical matrix which we call the quantum mechanical diffusion matrix;  from here  on we shall denote it  by $D$.  The diagonal elements of the latter along with the elements $D_{q_kp_k}$, corresponding to  each degree of freedom, have the same form as those associated with the one-dimensional harmonic oscillator. This is a direct result of the quadratic form of the Hamiltonian $H$. Hence, in our model, the  mutual interactions between the degrees of freedom do not affect the diagonal elements of the diffusion matrix. In particular,    the  {\it fluctuation-dissipation theorem} holds for every degree of freedom. Indeed, in the high-temperature limit, $k_B T\gg \hbar \omega_k$, after a Taylor expansion of the trigonometric function, we obtain  Einstein's  relation:
\begin{equation}
D_{p_kp_k}=\tilde \lambda_{kk} m_k k_B T\label{fldip},
\end{equation}  
where we have introduced the renormalized friction coefficient $\tilde\lambda_{kk}=\lambda_{kk}+\mu_{kk}$.

Furthermore it is quite interesting to notice that all the other diffusion coefficients in mixed coordinates and momenta~(\ref{co4})-(\ref{co6}), are the arithmetic mean of  two terms having the form of one of the coefficients~(\ref{co1})-(\ref{co3}) with, obviously, appropriate choice of both  the phenomenological constants and the coupling strengths. Conversely, the latter coefficients may be obtained from the most general ones~(\ref{co4})-(\ref{co6}), by setting $\omega_k=\omega_j$, $\mu_{jk}=\mu_{kj}=\mu_{kk}$, $\nu_{kj}=M_k\Omega_k^2$, $\kappa_{kj}=1/M_k$, and by observing that $\alpha_{kk}=\eta_{kk}=0$.  Notice, also, that  when $M_k=m_k$, and $\Omega_k=\omega_k$, then $D_{q_kp_k}\equiv 0$, a value which is usually used in nuclear  physics.

By analogy to equation~(\ref{fldip}), we may write, in the high temperature limit,
\begin{equation}
D_{p_kp_j}=\Lambda_{kj} \Bigl(\dfrac{m_k+m_j}{2}\Bigr) k_B T,\label{ein2}
\end{equation}
where the friction coefficient in this case is given by
\begin{equation}
\Lambda_{kj}=\frac{1}{m_k+m_j}\Bigl[(\lambda_{jk}+\mu_{jk})m_k+(\lambda_{kj}+\mu_{kj})m_j\Bigr].
\end{equation}
Equation~(\ref{ein2}) is nothing but Einstein's relation for a fictitious one-dimensional harmonic oscillator whose mass parameter is equal to $(m_k+m_j)/2$. When $m_k=m_j$, $\mu_{jk}=\mu_{kj}=0$, then $\Lambda_{kj}=(\lambda_{jk}+\lambda_{kj})/2$, i.e, the arithmetic mean of the friction coefficients $\lambda_{kj}$ and $\lambda_{jk}$. 

It is worth mentioning that in the linear response theory, the one-dimensional diffusion coefficient $D_{pp}$ is given in terms of the response function $\chi^{''}(t)$ of the operator that ensures the coupling of the degree of freedom to the heat bath by~\cite{hof1}
 \begin{equation}
 D_{pp}(\omega)=\coth\Bigl(\frac{\hbar\beta\omega}{2}\Bigl)\int_0^\infty dt i\chi^{''}(t)\sin(\omega t).
 \end{equation}
 Relation~(\ref{co5}) suggests that the multidimensional version of the above equation would be of the form
 \begin{equation}
 D_{p_kp_j}=\frac{1}{2}\Bigl[\coth\Bigl(\frac{\hbar\beta\omega_k}{2}\Bigl)\int_0^\infty dt i\chi_{jk}^{''}(t)\sin(\omega_k t)+\coth\Bigl(\frac{\hbar\beta\omega_j}{2}\Bigl)\int_0^\infty dt i\chi_{kj}^{''}(t)\sin(\omega_j t)\Bigr],
 \end{equation}
 where $\chi^{''}_{jk}$ is the response function corresponding to  the one-dimensional harmonic oscillator with frequency $\omega_k$ and mass $m_k$ which describes its coupling to both the heat reservoir and the oscillator with frequency $\omega_j$ and mass $m_j$.
\subsection{Constraints on the values of the transport coefficients}
Let us first begin  with briefly analyzing the conditions that should be satisfied by the coupling constants $\nu_{kj}$ and $\kappa_{kj}$ appearing in the expression of the Hamiltonian operator $\hat H$. In the special case where $N=2$, and $\mu_{kj}=\mu_{kk}=0$,  one can deduce, from simple mathematical considerations, that the kinetic  coupling strength satisfies the inequality
\begin{equation}
|\kappa_{1 2}|< \sqrt{\frac{1}{m_1 m_2}}. \label{inq1}
\end{equation}
 The coupling  constant $\nu_{12}$, on the other hand,  is such that
\begin{equation}
|\nu_{1 2}|<\sqrt{m_1 m_2 }\omega_1\omega_2. \label{inq2}
\end{equation} 
 Consider now the case $N=3$ with $m_3=m_2\neq m_1$, $\kappa_{12}=\kappa_{13}\neq\kappa_{23}$, $\nu_{12}=\nu_{13}\neq\nu_{23}$ and $\omega_2=\omega_3\neq\omega_1$. Then we should observe  the following conditions: \begin{equation}
|\kappa_{23}|<\frac{1}{m_2},\quad |\kappa_{12}|<\sqrt{\frac{1+m_2\kappa_{23}}{2 m_1m_2}}<\sqrt{\frac{1}{m_1m_2}},
\end{equation}
\begin{equation}
|\nu_{23}|< m_2\omega_2^2=m_3\omega_3^2, \quad |\nu_{12}|<\sqrt{\frac{1}{2}m_1\omega_1^2(\nu_{23}+m_2\omega_2^2)}<\sqrt{m_1 m_2}\omega_1\omega_2.
\end{equation}

The quantum character of the diffusion coefficients we have derived above may be perceived from the fundamental constraints they have to satisfy. Indeed, taking into account Cauchy-Schwartz inequality, we can infer from formulas~(\ref{coef1})-(\ref{coef2}) that
  \begin{eqnarray}
  &&D_{q_kq_k}D_{p_jp_j}-D_{q_kp_j}^2\ge\frac{\hbar^2}{4}\lambda_{kj}^2,\label{}\\
 && D_{q_kq_k} D_{q_jq_j}-D^2_{q_kq_j}\ge\frac{\hbar^2}{4} \alpha_{kj}^2,\\
 && D_{p_kp_k}D_{p_jp_j}-D_{p_kp_j}^2\ge\frac{\hbar^2}{4} \eta_{kj}^2, \\
 && \alpha_{kk}=\eta_{kk}=0.
  \end{eqnarray}
 These conditions ensure the non-negativity of the density matrix at any moment of time. There exist in the literature, however, other sets of diffusion coefficients which violate these constraints. This is the reason for which they are usually called {\it classical coefficients}~\cite{adam3} since a violation of the uncertainty relation may be observed at least at short times of the dynamics.

 In our multidimensional model, the friction coefficients cannot be freely chosen, in contrast to the one-dimensional case where the friction coefficient is dealt with as a free parameter which can be varied to reproduce the experimental data. As an illustration, consider the low temperature limit with $\mu_{kk}=\mu_{kj}=0$; then  it is a matter of algebra to show that
 \begin{equation}
 \sqrt{\lambda_{kk}\lambda_{jj}-\Bigl(\frac{m_k\omega_k}{m_j\omega_j}\Bigl)\lambda_{kj}^2}\ge\max\{\xi_{kj},\xi_{jk}\},
 \end{equation}
 where
 \begin{equation} \xi_{kj}=\frac{1}{2}\Biggl|\frac{\eta_{kj}+\nu_{kj}}{\sqrt{m_j m_k\omega_k\omega_j}}+\sqrt{m_km_j\omega_k\omega_k}(\alpha_{kj}-\kappa_{kj})\Biggl|.\end{equation}
  Furthermore, in case where $\lambda_{kj}=\lambda_{jk}=0$, then
 \begin{equation}
 |\alpha_{kj}|\le\sqrt{\frac{\lambda_{kk}\lambda_{jj}}{m_km_j\omega_k\omega_j}},\quad 
 |\eta_{kj}|\le\sqrt{\lambda_{kk}\lambda_{jj}{m_km_j\omega_k\omega_j}}, \quad k\ne j.
 \end{equation}
 The above conditions will be taken into account later in the numerical calculations. 
\section{Equations of motion\label{sec3}}
In what follows we shall be interested in the evolution in time of the mean values and variances of the coordinates and momenta operators. The latter may be calculated once the density matrix $\rho(t)$ is known. However, it is more convenient to work in the Heisenberg picture. Using formula~(\ref{mas3}), it can be shown that the evolution in time of any Heisenberg operator $\hat F$ is given by
\begin{eqnarray}
\frac{d\hat F}{dt}&=&\frac{i}{\hbar}[\hat H,\hat F]+\frac{1}{2\hbar^2}\sum_{kj}\Biggl\{(i\hbar\alpha_{kj}-2D_{q_kq_j})\Bigl(\{\hat F,\hat p_k\hat p_j\}-2\hat p_k \hat F\hat p_j\Bigl)+(i\hbar\eta_{kj}-2D_{p_kp_j})\nonumber\\&\times&\Bigl(\{\hat F,\hat q_k\hat q_j\}-2\hat q_k \hat F\hat q_j\Bigl)+(2D_{p_kq_j}+i\hbar\lambda_{jk})\Bigl(\{\hat F,\{\hat p_j, \hat q_k\}\}-2(\hat p_j \hat F \hat q_k+\hat q_k \hat F \hat p_j)\Bigl)\nonumber\\&-&2\hbar\lambda_{jk}\Bigl(\{\hat F,\hat p_j\hat q_k\}-2\hat q_k\hat F\hat p_j\Big)\Biggr\},\label{master}
\end{eqnarray}
where $\{\hat F,\hat G\}$ denotes the anticommuatator of the operators $\hat F$ and $\hat G$.
We recall here that the expectation values  and variances are explicitly defined as
\begin{eqnarray}
\sigma_{F}(t)&=&{\rm tr}(\rho \hat F(t)),\\ \sigma_{ F G}(t)&=&\frac{1}{2}{\rm tr}\Bigl(\rho\{\hat F(t),\hat G(t)\}\Bigl)-\sigma_{ F}(t)\sigma_{ G}(t),
\end{eqnarray}   
where ${\rm tr} (\hat F)$ denotes the trace of the operator $\hat F$.

  Let
\begin{equation}
\mathcal{V}(t)=\{\sigma_{q_1}(t),\sigma_{p_1}(t),\sigma_{q_2}(t),\sigma_{p_2}(t),\ldots\ldots ,\sigma_{ q_{N-1}}(t),\sigma_{p_{N-1}}(t),\sigma_{q_N}(t),\sigma_{p_N}(t)\}^{T},
\end{equation}
and
\begin{equation}
\sigma(t)=\begin{pmatrix}\sigma_{q_1q_1}(t)&\sigma_{q_1p_1}(t)&\sigma_{q_1q_2}(t)&\sigma_{q_1p_2}(t)&\cdots \cdots &\sigma_{q_1q_N}(t)&\sigma_{q_1p_N}(t)\\
\sigma_{p_1q_1}(t)&\sigma_{p_1p_1}(t)&\sigma_{p_1q_2}(t)&\sigma_{p_1p_2}(t)&\cdots \cdots&\sigma_{p_1q_N}(t)&\sigma_{p_1p_N}(t)\\
\vdots&\vdots&  \vdots&\vdots &\cdots \cdots& \vdots&\vdots\\\vdots&\vdots&  \vdots&\vdots &\cdots \cdots& \vdots&\vdots\\\sigma_{q_Nq_1}(t)&\sigma_{q_Np_1}(t)&\sigma_{q_N q_2}(t)&\sigma_{q_Np_2}(t)&\cdots \cdots &\sigma_{q_Nq_N}(t)&\sigma_{q_Np_N}(t)\\
\sigma_{p_Nq_1}(t)&\sigma_{p_Np_1}(t)&\sigma_{p_N q_2}(t)&\sigma_{p_Np_2}(t)&\cdots \cdots&\sigma_{p_Nq_N}(t)&\sigma_{p_Np_N}(t)\\
\end{pmatrix}
\end{equation}
Then, by virtue of equation~(\ref{master}),  one can show that 
\begin{eqnarray}
&&\frac{d\mathcal{V}(t)}{dt}=M\mathcal{V}(t),\label{evol1}\\
&&\frac{d\sigma(t)}{dt}=M\sigma(t)+\sigma(t)M^T+2D,\label{evol2}
 \end{eqnarray}
where 
 \begin{eqnarray} 
 M= {\footnotesize \begin{pmatrix}
        -\lambda_{11}+\mu_{11}&\tfrac{1}{m1} & -\lambda_{12}+\mu_{12}&-\alpha_{12}+\kappa_{12} &\cdots  &  \cdots &-\lambda_{1N}+\mu_{1N}& -\alpha_{1N}+\kappa_{1N} \\
        - m_1 \omega_1^2 &-\lambda_{11}-\mu_{11}&\eta_{12}-\nu_{12}&-\lambda_{21}-\mu_{21}& \cdots &  \cdots &\eta_{1N}-\nu_{1N}&-\lambda_{N1}-\mu_{N1}\\
        -\lambda_{21}+\mu_{21}&\alpha_{12}+\kappa_{12}&-\lambda_{22}+\mu_{22}&\tfrac{1}{m_2} & \cdots & \cdots & -\lambda_{2N}+\mu_{2N} &-\alpha_{2N}+\kappa_{2N} \\
       -\eta_{12}-\nu_{12}&-\lambda_{12}-\mu_{12}& - m_2 \omega_2^2 &-\lambda_{22}-\mu_{22}& \cdots &  \cdots &\eta_{2N}-\nu_{2N} &-\lambda_{N2}-\mu_{N2}\\ 
       \vdots & \vdots & \vdots & \vdots &\vdots &\vdots &\vdots & \vdots\\
       \vdots & \vdots & \vdots & \vdots & \vdots &   \vdots & \vdots & \vdots\\
       -\lambda_{N1}+\mu_{N1} &\alpha_{1N}+\kappa_{1N}&-\lambda_{N2}+\mu_{N2} &  \cdots & \cdots & \cdots &-\lambda_{NN}+\mu_{NN}& \tfrac{1}{m_N}\\
       -\eta_{1N}-\nu_{1N}& -\lambda_{1N}-\mu_{1N}&-\eta_{2N}-\nu_{2N}&  \cdots & \cdots & \cdots &-m_{N}\omega_N^2 & -\lambda_{NN}-\mu_{NN}
        \label{matr}
        \end{pmatrix}},
\end{eqnarray}
 and $D$ is the diffusion matrix:
 \begin{equation}
D=\begin{pmatrix}D_{q_1q_1}&D_{q_1p_1}&D_{q_1q_2}&D_{q_1p_2}&\cdots \cdots &D_{q_1q_N}&D_{q_1p_N}\\
D_{p_1q_1}&D_{p_1p_1}&D_{p_1q_2}&D_{p_1p_2}&\cdots \cdots&D_{p_1q_N}&D_{p_1p_N}\\
\vdots&\vdots&  \vdots&\vdots &\cdots \cdots& \vdots&\vdots\\\vdots&\vdots&  \vdots&\vdots &\cdots \cdots& \vdots&\vdots\\D_{q_Nq_1}&D_{q_Np_1}&D_{q_N q_2}&D_{q_Np_2}&\cdots \cdots &D_{q_Nq_N}&D_{q_Np_N}\\
D_{p_Nq_1}&D_{p_Np_1}&D_{p_N q_2}&D_{p_Np_2}&\cdots \cdots&D_{p_Nq_N}&D_{p_Np_N}\\
\end{pmatrix}.
\end{equation}
The solution of equation~(\ref{evol1}) is simply
\begin{equation}
\mathcal V(t)=\exp(M t)\mathcal V(0),
\end{equation}
  whereas that corresponding to equation~(\ref{evol2}) may be obtained by noting the following property:
  \begin{equation}
  \frac{d}{dt}\Bigl\{e^{A t} B e^{ C t}\Bigr \}= A e^{A t} B e^{ C t}+e^{A t} B e^{ C t} C, \end{equation}
 where the  matrices $A$, $B$ and $C$ do not depend on time.  Consequently, the time development of the matrix $\sigma$ is given by
  \begin{equation}
  \sigma(t)=\exp(M t)(\sigma(0)-\tilde\sigma)\exp(M t)^T+\tilde\sigma,
  \end{equation}
  where the matrix $\tilde\sigma$ satisfies 
  \begin{equation}
  M\tilde\sigma+\tilde\sigma M^T+2D=0.
  \end{equation}
 In general  the latter equation yields a set of $\frac{2N}{2}(2N+1)=N(2N+1)$ algebraic linear equations,  the unknowns of which are the elements of the matrix $\tilde\sigma$. 
 
 In our investigation we are assuming that the asymptotic sate of the system is the Gibbs state~(\ref{gibbs}).  Under this condition it can be verified that the expectation values and variances tend to  
 \begin{eqnarray}
 \lim_{t\to\infty}\sigma_{q_k}(t)&=& \lim_{t\to\infty}\sigma_{p_k}(t)=\lim_{t\to\infty}\sigma_{p_kq_k}(t)=0,\label{asy1}\\
 \lim_{t\to\infty}\sigma_{q_kq_k}(t)&=&\sigma_{q_kq_k}(\infty)=\frac{\hbar}{2m_k\omega_k}\coth\frac{\hbar\beta\omega_k}{2},\label{asy2}\\
 \lim_{t\to\infty}\sigma_{p_kp_k}(t)&=&\sigma_{p_kp_k}(\infty)=\frac{\hbar}{2} m_k\omega_k\coth\frac{\hbar\beta\omega_k}{2},\label{asy3}\\
 \lim_{t\to\infty}\sigma_{q_kq_j}(t)&=&\lim_{t\to\infty}\sigma_{p_k p_j}(t)=\lim_{t\to\infty}\sigma_{p_kq_j}(t)=0,\quad k\ne j.\label{asy4}
 \end{eqnarray}
Hence it is possible to link the diffusion coefficients to the asymptotic variances by simple expressions. We have, for instance,
 \begin{eqnarray}
D_{q_kp_j}=D_{p_jq_k}=\frac{1}{2}\Bigl[(\eta_{kj}+\nu_{kj})\sigma_{q_kq_k}(\infty)+(\alpha_{kj}-\kappa_{kj})\sigma_{p_jp_j}(\infty)\Bigl].
\end{eqnarray}

 Some remarks are in order here. First of all, the fact that the asymptotic expectation values  $\sigma_{q_k}(\infty)$ and $\sigma_{p_k}(\infty)$ are zero implies that  the real part of all the eigenvalues of the matrix $M$ should be negative, a fact that is equivalent to  the condition $\exp(M t)\to0$ as $t\to\infty$. This, actually, imposes further conditions on the relevant parameters of the model. In particular we find that the matrix  $\tilde\sigma$ is nothing but the asymptotic variance matrix, that is, $\tilde\sigma=\sigma(\infty)$.
 
   Notice also that the generalized Heisenberg uncertainty relation
 \begin{equation}
 \sigma_{q_kq_k}(t)\sigma_{p_kp_k}(t)-\sigma_{q_kp_k}(t)^2\ge\frac{\hbar^2}{4}\label{heis}
 \end{equation}
 should be observed,  since the operators $\hat p(t)$ and $\hat q(t)$ satisfy the usual canonical commutation relation at any moment of the time. It has been shown, however, that when the fundamental constraints imposed on the diffusion coefficients are not satisfied,  then it may happen that the inequality~(\ref{heis}) is violated at certain interval of the time.

  The expectation value of the Hamiltonian $\hat H$ can be calculated using equation~(\ref{master}). The resulting formula is quite cumbersome, and we shall not display it here. Nevertheless, by direct calculation one can verify that
 \begin{eqnarray}
 E=\lim_{t\to\infty }{\rm tr}(\rho \hat H(t))=\frac{\hbar}{2}\sum_k^N\omega_k\coth\frac{\hbar\beta\omega_k}{2}.
 \end{eqnarray}

\section{Application to heavy-ion collisions\label{sec4}}
In what follows, we shall apply the results obtained above to the description of the motion of a dinuclear system (DNS) in the  charge and mass asymmetry coordinates
\begin{equation}
\eta_Z=\frac{Z_1-Z_2}{Z_1+Z_2}, \qquad \eta_N=\frac{N_1-N_2}{N_1+N_2}.\end{equation} 
 Here $Z_1$, $N_1$ and $Z_2$, $N_2$ are, respectively, the charge number and neutron number of the nuclei.   The advantages of the DNS come into play in the description of the various reaction channels in heavy ions collisions, such as  fission and fusion of atomic nuclei.
 
Based on the work of Hahn et {\it al}~\cite{fast}, Sandulescu {\it et al}~\cite{exp1}, the authors of~\cite{exp2} proposed an analytically solvable quantum-mechanical model describing the charge and mass distribution in heavy-ion collisions. There, the investigation consists in solving the following time-depending Schr\"{o}dinger  equation:
\begin{eqnarray}
\Biggl[-\frac{\hbar^2}{2M_{ZZ}}\frac{\partial}{\partial\eta_Z^2}-\frac{\hbar^2}{2M_{NN}}\frac{\partial}{\partial\eta_N^2}+\frac{1}{2} k_Z\eta_Z^2+\frac{1}{2}k_N \eta_N^2&-&k_{ZN}\eta_N\eta_Z\Biggl]\psi(\eta_Z,\eta_N,t)\nonumber\\&=&i\hbar\frac{\partial}{\partial t}\psi(\eta_Z,\eta_N,t),\label{shro}
\end{eqnarray}
where $M_{ZZ}$ and $M_{NN}$ are mass  parameters, $k_Z$ and $k_N$ are stiffness parameters, and $k_{ZN}$ is the coupling constant. They have, however, neglected dissipation by assuming that the mass and charge asymmetry degrees of freedom are isolated from the other collective and intrinsic degrees of freedom. They also assumed that the neutron and proton mobilities are uncorrelated; this is the reason why there is no momentum-momentum coupling in the above equation. Before we proceed further, note that the quadratic form of the Hamiltonian in (\ref{shro}), valid only for nearly grazing collisions,  was obtained from  an expansion around the energy surface  minimum $\eta_N=\eta_Z=0$ of the  potential energy between two colliding nuclei, which is defined as the sum of the usual liquid-drop energy, the Coulomb contribution due to charged protons, the rotational energy (proportional to the square of the total angular momentum), and the proximity nuclear potential. The last two contributions depend strongly on the relative distance between the two interaction partners. 

Our task here is to investigate the effect of the transport coefficients  on the dynamics of the compound nuclear system by introducing through Lindblad's formalism damping effects . Attention will be given to the influence of the coupling between the collective degrees of freedom on the evolution in time of the expectation values and variances (see reference~\cite{two1} for further discussion).  
\begin{figure}[h]
{\centering{
\resizebox*{0.45\textwidth}{!}{\includegraphics{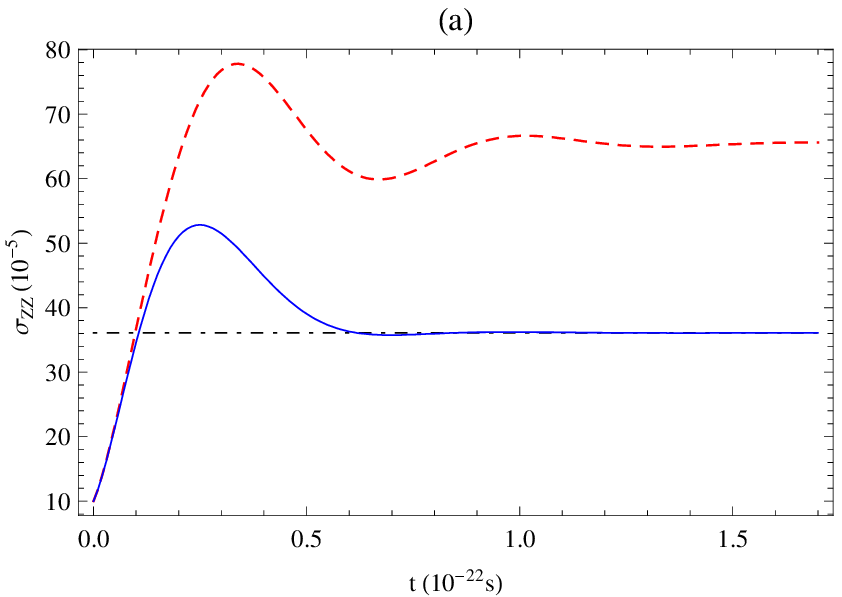}}
\resizebox*{0.46\textwidth}{!}{\includegraphics{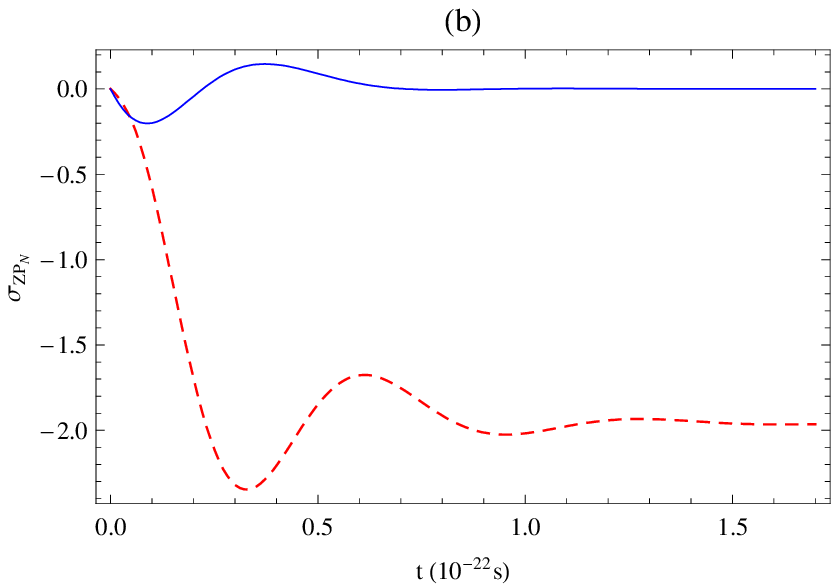}}
\par}}

\caption{\label{fig1} (Color online) Time evolution of: (a) $\sigma_{ZZ}(t)$,  and (b) $\sigma_{Z p_N}(t)$ 
for zero off-diagonal elements of $D$ (dashed lines) and nonzero off-diagonal elements of $D$ (solid lines). The dot-dashed line represents the asymptotic value corresponding  to the Gibbs state. The parameters are $M_{ZZ}=M_{NN}= 461.6344 \hbar^2/{\rm MeV}$, $\hbar\omega_Z=2.9468${\rm MeV}, $\hbar\omega_N=2.9288$ {\rm MeV}, $\nu_{ZN}=-1869$ {\rm  MeV}, $\hbar \lambda_{ZZ}=\hbar\lambda_{NN}=2$ {\rm MeV}, $T=5$ {\rm MeV}, $\sigma_{ZZ}(0)=10^{-4}$, $\sigma_{p_Zp_Z}(0)=\hbar^2/(4\sigma_{ZZ}(0))$, $\sigma_{NN}=10^{-3}$, $\sigma_{p_Np_N}(0)=\hbar^2/(4\sigma_{NN}(0))$; all other parameters are set to zero.}
\end{figure}
\begin{figure}[htba]
{\centering{
\resizebox*{0.40\textwidth}{!}{\includegraphics{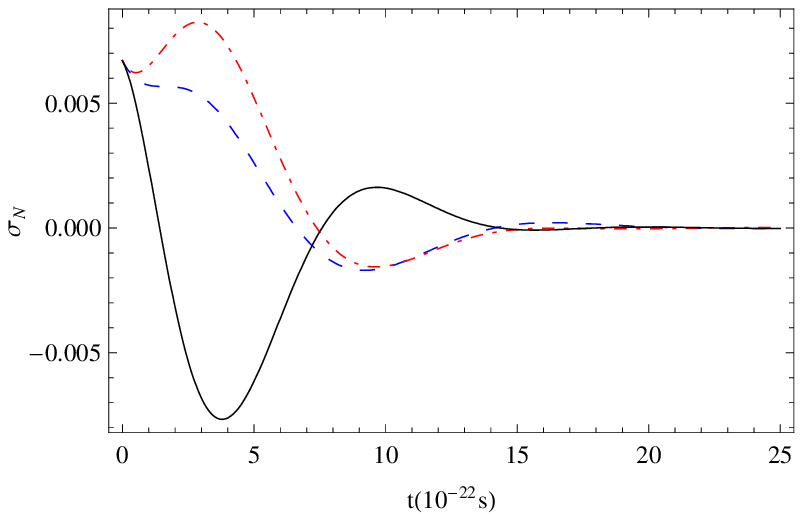}}
\resizebox*{0.40\textwidth}{!}{\includegraphics{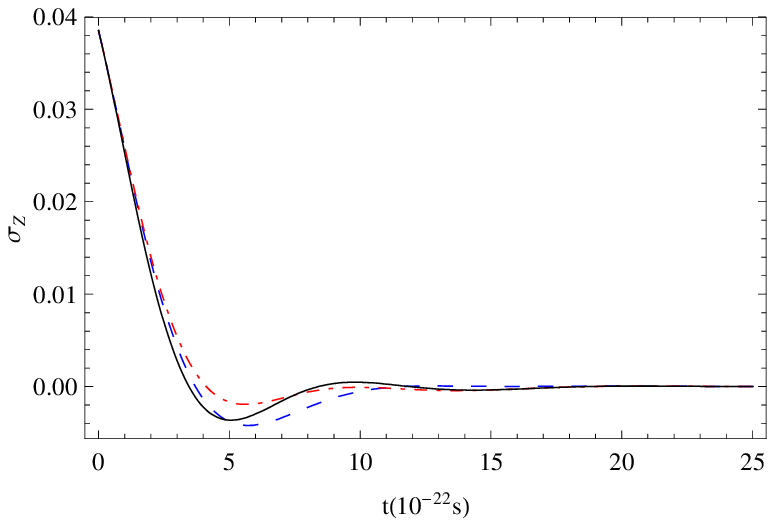}}
\resizebox*{0.40\textwidth}{!}{\includegraphics{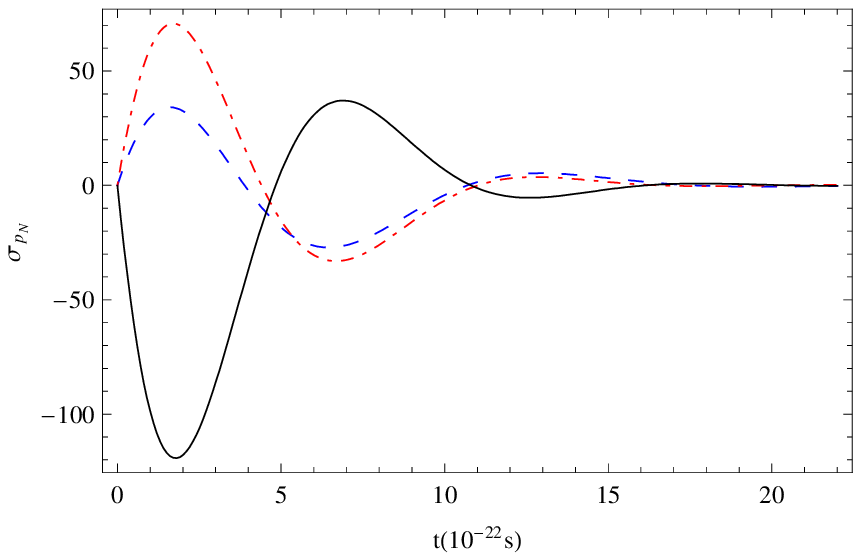}}
\resizebox*{0.40\textwidth}{!}{\includegraphics{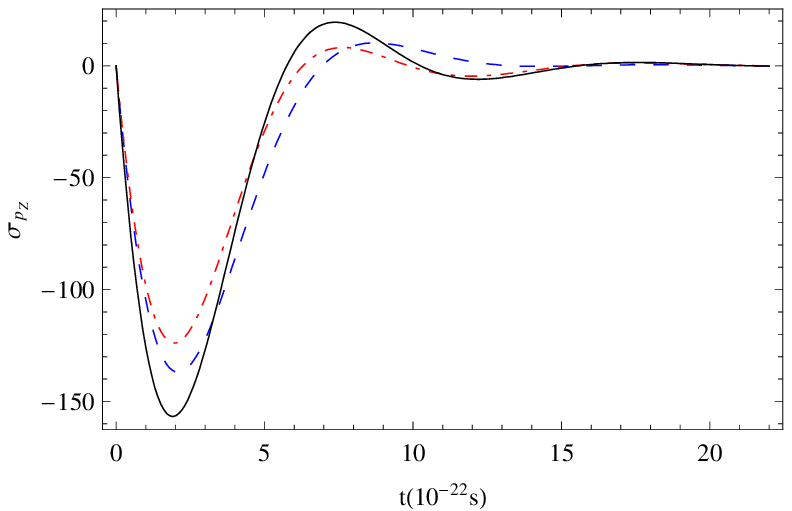}}
\par}}

\caption{\label{fig2} (Color online) Time evolution of the expectation values  $\sigma_N(t)$, $\sigma_Z(t)$, $\sigma_{p_N}(t)$ and $\sigma_{p_Z}(t)$
for different values of the coupling  constant $\nu_{NZ}$; $\nu_{NZ}=3000$ {\rm MeV} (solid line), $\nu_{NZ}=-1968$ {\rm MeV} (dashed line) and $\nu_{NZ}=-3000$ {\rm MeV} (dot-dashed line). Here $T=2$ {\rm MeV}, $\sigma_{p_Z}(0)=\sigma_{p_N}(0)=0$; the other parameters are the same as figure~\ref{fig1}.}
\end{figure}

\subsection{Illustrative calculations} 
Unless otherwise stated, the model parameters we shall use in the sequel are those corresponding to the system ${\rm ^{129}Xe+{\rm ^{124}{Sn}}}$. The  stiffness parameters $k_Z$ and $k_N$ are found to be equal to $4009$ {\rm MeV} and $3960$ {\rm MeV}, respectively~\cite{exp2}. The average mass parameters, calculated within the framework of the hydrodynamical theories, are given by $M_{ZZ}=M_{NN}\approx 461.6344 \hbar^2/{\rm MeV}$. This corresponds to a value of the angular frequencies of $\hbar\omega_Z\approx2.9468$ {\rm MeV}, $\hbar\omega_N\approx2.9288$ {\rm MeV}. The calculation gives a value of $3739$ {\rm MeV} for the coupling constant $k_{ZN}$, which implies that $\nu_{ZN}=\nu_{NZ}=-1869$ {\rm  MeV}. It can easily be checked that these values satisfy the condition (\ref{inq2}). The friction coefficient $\lambda_{kk}$ (with $k\equiv N, Z$) has the dimension of the angular frequency $\omega_k$; they are, in general, of the same order. More precisely, due to fast charge equilibration, we should have $2\lambda_{NN}> \omega_N$ and $2\lambda_{ZZ}> \omega_Z$. The initial value of the mass and charge asymmetries can easily be calculated;  one can find that $\eta_Z(0)=0.0385$, and $\eta_N=0.0067$.

 \begin{figure}[htba]
{\centering{
\resizebox*{0.50\textwidth}{!}{\includegraphics{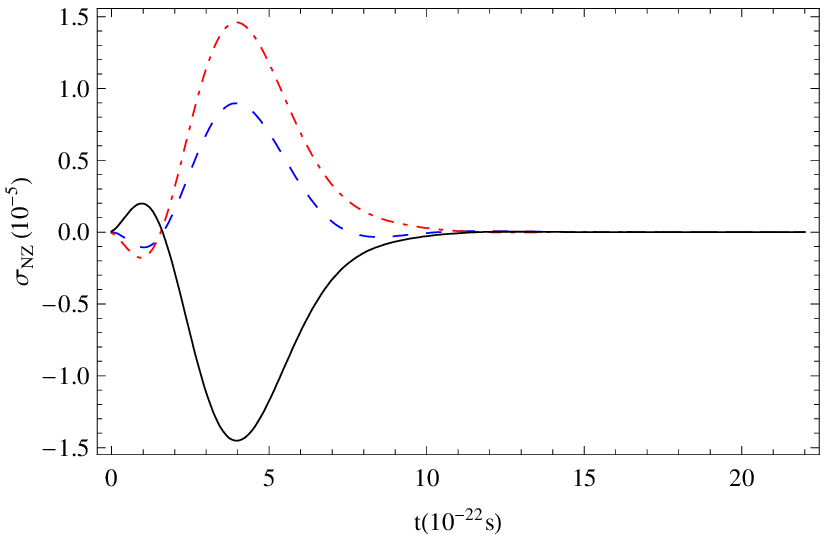}}
\par}}

\caption{\label{fig3} (Color online)  Evolution in time of the variance  $\sigma_{NZ}(t)$ for different values of the coupling  constant $\nu_{NZ}$; $\nu_{NZ}=3000$ {\rm MeV} (solid line), $\nu_{NZ}=-1968$ {\rm MeV} (dashed line) and $\nu_{NZ}=-3000$ {\rm MeV} (dot-dashed line). Here $T=2$ {\rm MeV}; the other parameters are the same as figure~\ref{fig1}.}
\end{figure}
\begin{figure}[htba]
{\centering{
\resizebox*{0.40\textwidth}{!}{\includegraphics{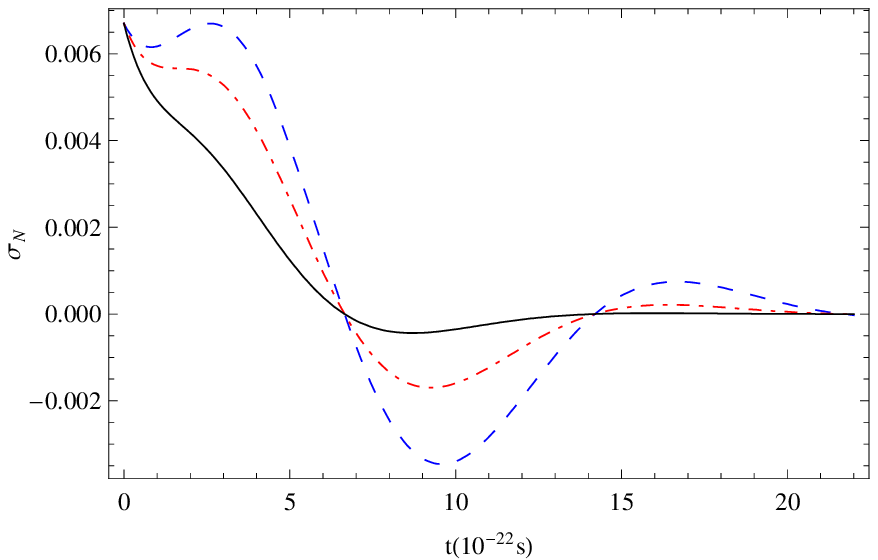}}
\resizebox*{0.40\textwidth}{!}{\includegraphics{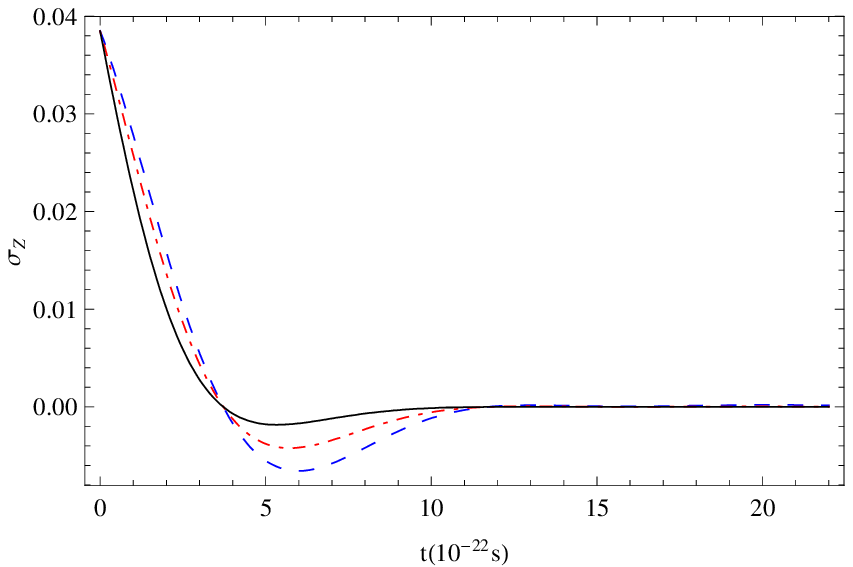}}
\resizebox*{0.40\textwidth}{!}{\includegraphics{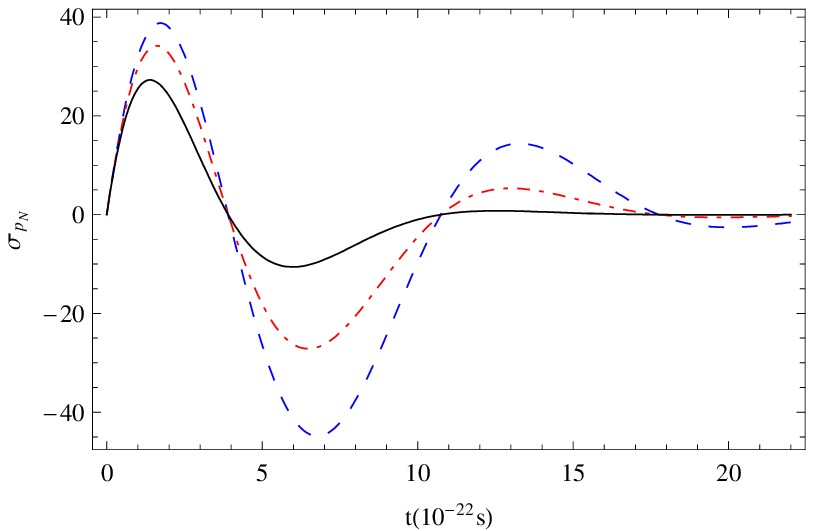}}
\resizebox*{0.40\textwidth}{!}{\includegraphics{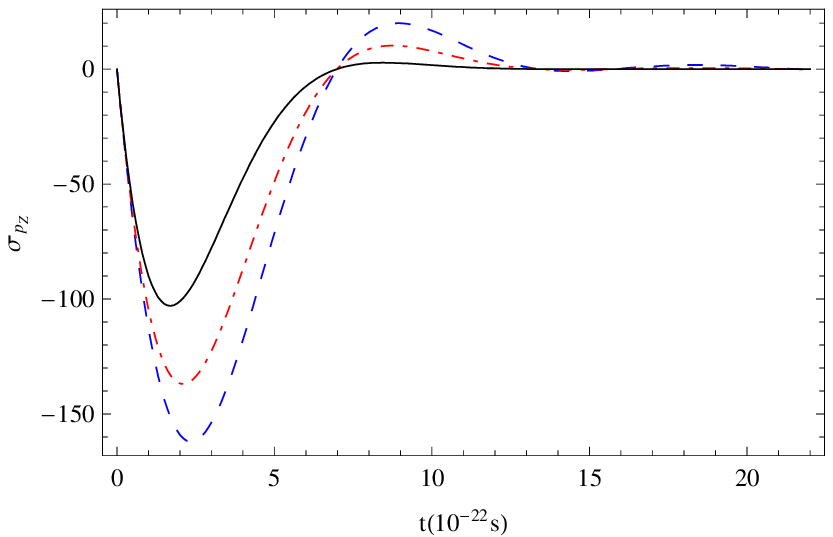}}
\par}}

\caption{\label{fig4} (Color online) Time evolution of the expectation values  $\sigma_N(t)$, $\sigma_Z(t)$, $\sigma_{p_N}(t)$ and $\sigma_{p_Z}(t)$
for different values of the friction coefficients $\lambda_{NN}$ and $\lambda_{ZZ}$; $\hbar\lambda_{NN}=\hbar\lambda_{ZZ}=3$ {\rm MeV} (solid line), $\hbar\lambda_{NN}=\hbar\lambda_{ZZ}=2$ {\rm MeV} (dot-dashed line) and $\hbar\lambda_{NN}=\hbar\lambda_{ZZ}=1.6$ {\rm MeV} (dashed line). Here $T=2$ {\rm MeV}, $\nu_{ZN}=-1869$ {\rm MeV}, $\sigma_{p_Z}(0)=\sigma_{p_N}(0)=0$; the other parameters are the same as figure~\ref{fig1}.}
\end{figure}
\begin{figure}[htba]
{\centering{
\resizebox*{0.50\textwidth}{!}{\includegraphics{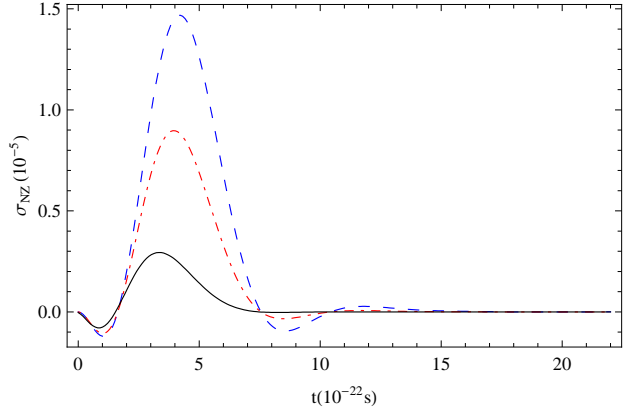}}
\par}}

\caption{\label{fig5} (Color online)  Evolution in time of the variance  $\sigma_{NZ}(t)$ for different values of the friction coefficients $\lambda_{NN}$ and $\lambda_{ZZ}$; $\hbar\lambda_{NN}=\hbar\lambda_{ZZ}=3$ {\rm MeV} (solid line), $\hbar\lambda_{NN}=\hbar\lambda_{ZZ}=2$ {\rm MeV} (dot-dashed line) and $\hbar\lambda_{NN}=\hbar\lambda_{ZZ}=1.6$ {\rm MeV} (dashed line). Here $T=2$ {\rm MeV}; the other parameters are the same as figure~\ref{fig1}.}
\end{figure}

\begin{figure}[htba]
{\centering{
\resizebox*{0.40\textwidth}{!}{\includegraphics{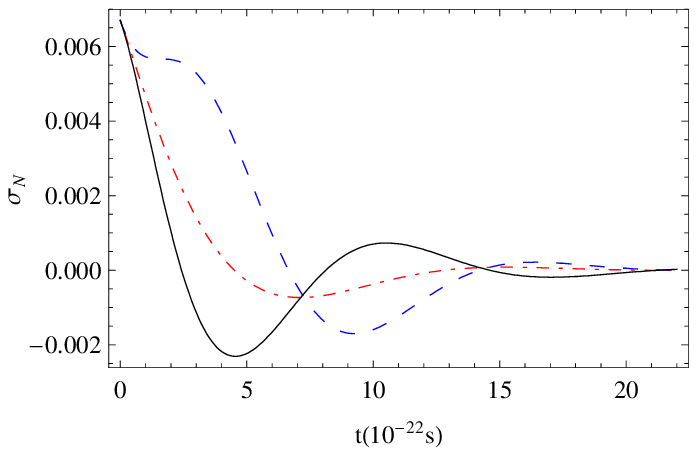}}
\resizebox*{0.40\textwidth}{!}{\includegraphics{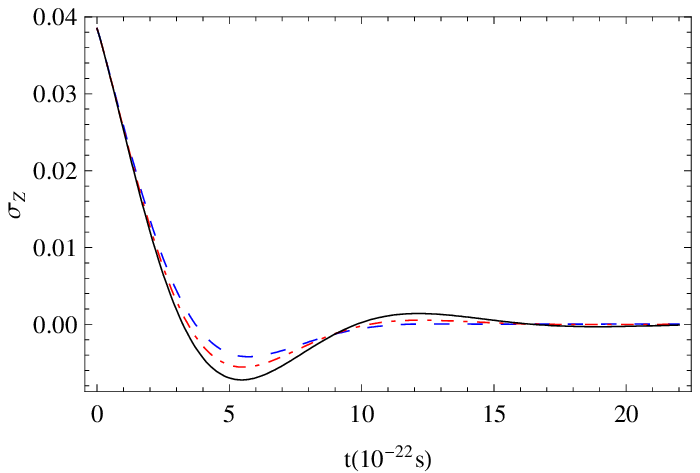}}
\resizebox*{0.40\textwidth}{!}{\includegraphics{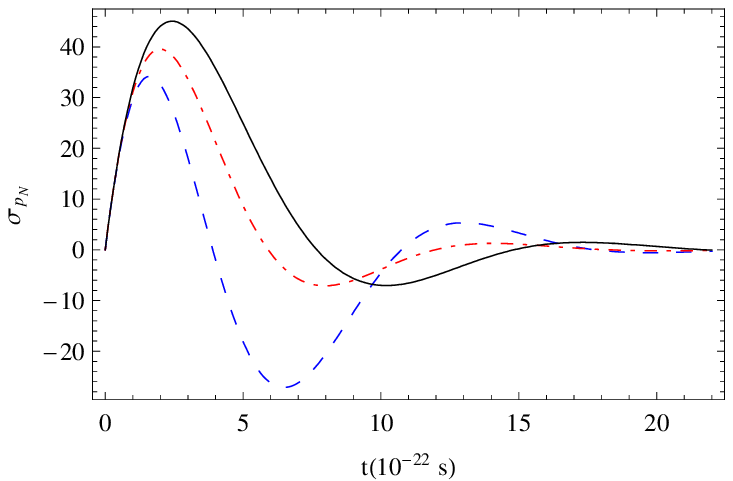}}
\resizebox*{0.40\textwidth}{!}{\includegraphics{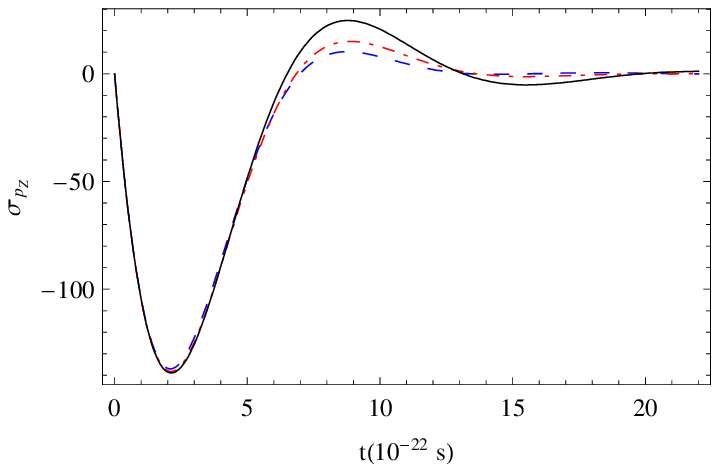}}
\par}}

\caption{\label{fig6} (Color online) Time evolution of the expectation values  $\sigma_N(t)$, $\sigma_Z(t)$, $\sigma_{p_N}(t)$ and $\sigma_{p_Z}(t)$
for different values of the coupling  constant $\kappa_{NZ}$; $\kappa_{NZ}=33\times 10^{38}\ {\rm MeV^{-1}}\ {\rm s}^{-2}$ (solid line), $\kappa_{NZ}=20\times 10^{38}\ {\rm MeV^{-1}}\ {\rm s}^{-2}$ (dot-dashed line) and $\kappa_{NZ}=0$ (dashed line). Here $T=2$ {\rm MeV}, $\sigma_{p_Z}(0)=\sigma_{p_N}(0)=0$; the other parameters are the same as figure~\ref{fig1}.}
\end{figure}
In order to illustrate the importance of the off-diagonal elements of the diffusion matrix we display  in figure~\ref{fig1} the evolution in time of the variances  $\sigma_{ZZ}(t)$ and $\sigma_{Zp_N}(t)$ for both zero and nonzero off-diagonal elements. We see that the asymptotic values of the variances when the off-diagonal elements are set to zero do not correspond to the Gibbs state~(\ref{gibbs}) as indicated by the dot-dashed line in the above figure. Therefore, we conclude that the  behaviour of the dynamics of the compound nuclear system  depend strongly on the values of the diffusion coefficients.  
 \begin{figure}[htba]
{\centering{
\resizebox*{0.50\textwidth}{!}{\includegraphics{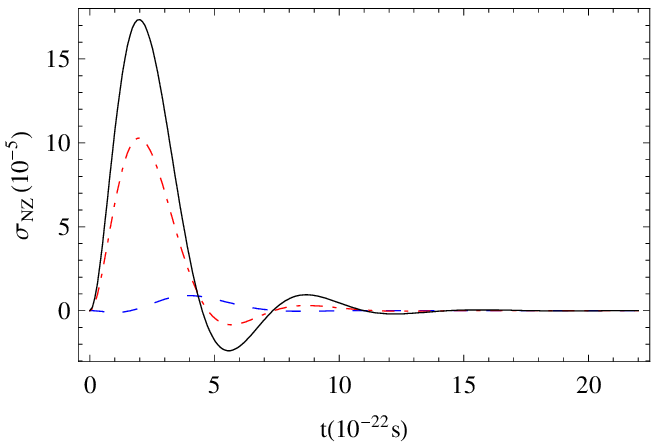}}
\par}}

\caption{\label{fig7} (Color online)  Evolution in time of the variance  $\sigma_{NZ}(t)$ for different values of the coupling constant $\kappa_{NZ}$; $\kappa_{NZ}=33\times 10^{38}\ {\rm MeV^{-1}}\ {\rm s}^{-2}$ (solid line), $\kappa_{NZ}=20\times 10^{38}\ {\rm MeV^{-1}}\ {\rm s}^{-2}$ (dot-dashed line) and $\kappa_{NZ}=0$ (dashed line). Here $T=2 $ {\rm MeV}, $\sigma_{p_Z}(0)=\sigma_{p_N}(0)=0$; the other parameters are the same as figure~\ref{fig1}.}
\end{figure}

 An example of  the development in time of the expectation values and variances  of the charge and neutron asymmetry coordinates for different values of the coupling strength $\nu_{NZ}$ is displayed in figures~\ref{fig2} and~\ref{fig3}. It can easily be seen that the expectation values decay faster for large positive values of the latter parameter. The decay gets slower as we decrease $\nu_{NZ}$ to negative values, which implies that the evolution of the mean values depends on the sign of the coupling constant. This  fact  manifests itself even though the system is nearly symmetric. The differences in the evolution in time of the proton and neutron asymmetry coordinates may be explains by the fact that the dynamics is sensitive to the initial expectation values.  Notice that the situation is, however, slightly different regarding the evolution in time of the variances $\sigma_{NZ}$ and $\sigma_{p_Np_Z}$ (their values quantify  the correlation between the two degrees of freedom).  In this case the curves corresponding to coupling constants having the same magnitude but with different signs  are symmetrical, see figure~\ref{fig3}. The investigation shows that the above result does not hold  when  the initial values of $\sigma_{NZ}$ and $\sigma_{p_Np_Z}$ are different from zero. The other variances do not  change much when varying $\nu_{NZ}$.

Let us now investigate the effect of the friction coefficients and the kinetic coupling constant $\kappa_{NZ}$ on the behaviour of the DNS. Figure~\ref{fig4} displays the time dependence of the expectation values for different values of $\lambda_{NN}$ and $\lambda_{ZZ}$. As expected we see that the decay of the above quantities is less appreciable for small values of the friction coefficients. This result does not qualitatively change with nonzero values of the remaining model parameters. It is also found that except $\sigma_{NZ}$ (see figure~\ref{fig5}), and $\sigma_{p_Np_Z}$, the other variances are not much affected by changing the values of the friction coefficients. The time development of the centroids for different values of $\kappa_{NZ}$ is illustrated in figure~\ref{fig6}. Once again we find that the decay is faster for large values of the coupling constant, whereas the correlation between the proton and neutron asymmetry coordinates becomes larger, as shown in figure~\ref{fig7}. When the initial values of $\sigma_{NZ}$ and $\sigma_{p_Np_Z}$ are zero then the above quantities are symmetrical with respect to the change of the sign of $\kappa_{NZ}$ (see figure~\ref{fig3} for a similar situation). The other variances are robust with regard to  the variation of the latter constant.

\subsection{Comparison with experimental data}
\begin{figure}[h]
{\centering{
\resizebox*{0.48\textwidth}{!}{\includegraphics{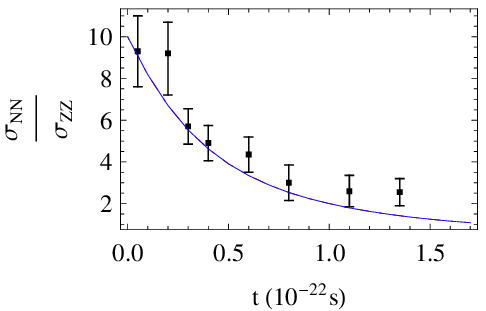}}
\resizebox*{0.46\textwidth}{!}{\includegraphics{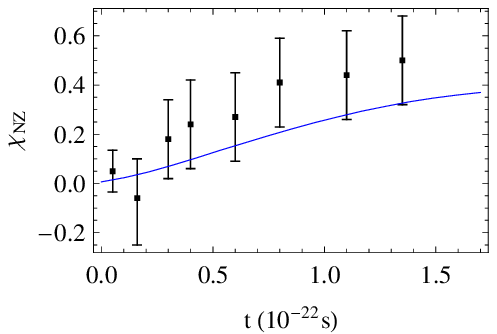}}
\par}}

\caption{\label{fig8} (Color online) Experimental data~\cite{exp3} (dots) along with the theoretical  curves (solid lines) corresponding to the proton to neutron variances ratio (left) and the correlation coefficient $\chi_{NZ}(t)$ (right) as functions of time for the reaction ${\rm ^{129}Xe+{\rm ^{124}{Sn}}}$.  Here  the parameters are: $M_{ZZ}=M_{NN}= 461.6344 \hbar^2/{\rm MeV}$, $\hbar\omega_Z=2.9468$\ {\rm MeV}, $\hbar\omega_N=2.9288$ {\rm MeV}, $\nu_{ZN}=-1869$ {\rm  MeV}, $\hbar \lambda_{ZZ}=\hbar\lambda_{NN}=2$ {\rm MeV}, $T=0.02$ {\rm MeV}, $\sigma_{ZZ}(0)=10^{-4}$, $\sigma_{p_Zp_Z}(0)=\hbar^2/(4\sigma_{ZZ}(0))$, $\sigma_{NN}=10^{-3}$, $\sigma_{p_Np_N}(0)=\hbar^2/(4\sigma_{NN}(0))$, $\sigma_{NZ}(0)=0$, $\alpha_{ZN}=33\times 10^{38} \ {\rm MeV^{-1}\ s^{-2}}$; all other parameters are set to zero.}
\end{figure}

Now we are going to assess  the results of our  model by comparing them with  the experimental data obtained by Sch\"ull {\it et al}~\cite{exp3}. For this reason we have calculated the ratio of neutron to proton variances and the correlation coefficient $\chi_{NZ}$, defined by  
\begin{equation}
\chi_{NZ}(t)=\frac{\sigma_{NZ}(t)}{\sqrt{\sigma_{NN}(t)\sigma_{ZZ}(t)}},
\end{equation}
for the reaction ${\rm ^{129}Xe+{\rm ^{124}{Sn}}}$. The results are displayed in figure~\ref{fig8}. The theoretical curves (solid lines) were obtained for a value of the friction coefficients $\lambda_{NN}=\lambda_{ZZ}=2\ {\rm MeV}/\hbar$, with $\alpha_{ZN}=-\alpha_{NZ}= 33\times 10^{38} \ {\rm MeV^{-1}\ s^{-2}}$ which gives the best fit to  the experimental data without violating the fundamental constraints on the transport coefficients. We can see that with the diffusion coefficients~(\ref{co1})-(\ref{co6}), the theoretical values of the ratio $\sigma_{NN}(t)/\sigma_{ZZ}(t)$ are in quite good agreement with the experimental ones in the interval $0\le t \leq 10^{-22}\ {\rm s}$. On the other hand, though not in perfect agreement with the experimental outcomes, the model gives improved results for the correlation coefficient as compared with those of~\cite{exp2} and~\cite{exp4}. This difference may be explained by the nonzero value of the parameter $\alpha_{ZN}$, which is responsible for the creation of  momentum-momentum correlations between the charge and mass asymmetry coordinates, even though  the motion of the neutrons and protons was assumed to be  initially uncorrelated ($\kappa_{NZ}=0$). We have further checked the validity of the above results by coupling the neutron and proton asymmetry coordinates to the relative motion of the nuclei. It turns out that the only difference between the two cases resides in a small diminution of the value of the friction coefficients. It is worth mentioning that the ratio of neutron to proton variances does not depend much on the value of $\alpha_{NZ}$. Also,  nonzero initial values of $\sigma_{NZ}$ do not significantly improve $\chi_{NZ}(t)$.    
\subsection{Penetration enhancement due to dissipation in sub-barrier processes} As a second application, in connection with the description of the fusion process, we now investigate the penetration of Gaussian wave packets through  a potential barrier approximated by a two-dimensional inverse harmonic oscillator. It should be stressed  that there exist no metastable states for this kind of potentials; a more appropriate one would be composed of a potential well smoothly linked to a parabolic barrier.

 \begin{figure}[htba]
{\centering{
\resizebox*{0.40\textwidth}{!}{\includegraphics{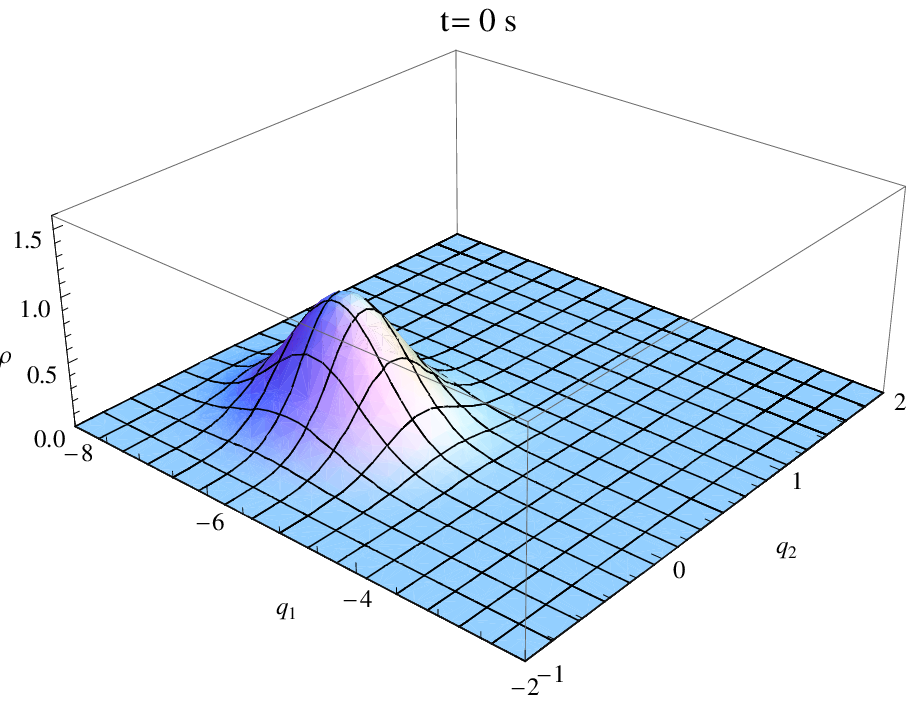}}
\resizebox*{0.40\textwidth}{!}{\includegraphics{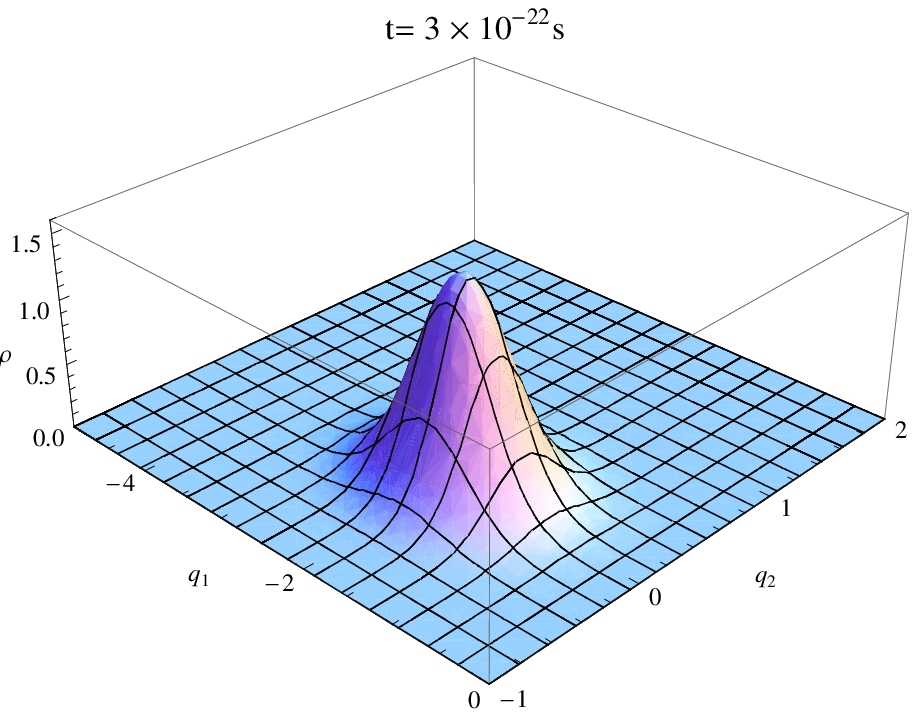}}
\resizebox*{0.40\textwidth}{!}{\includegraphics{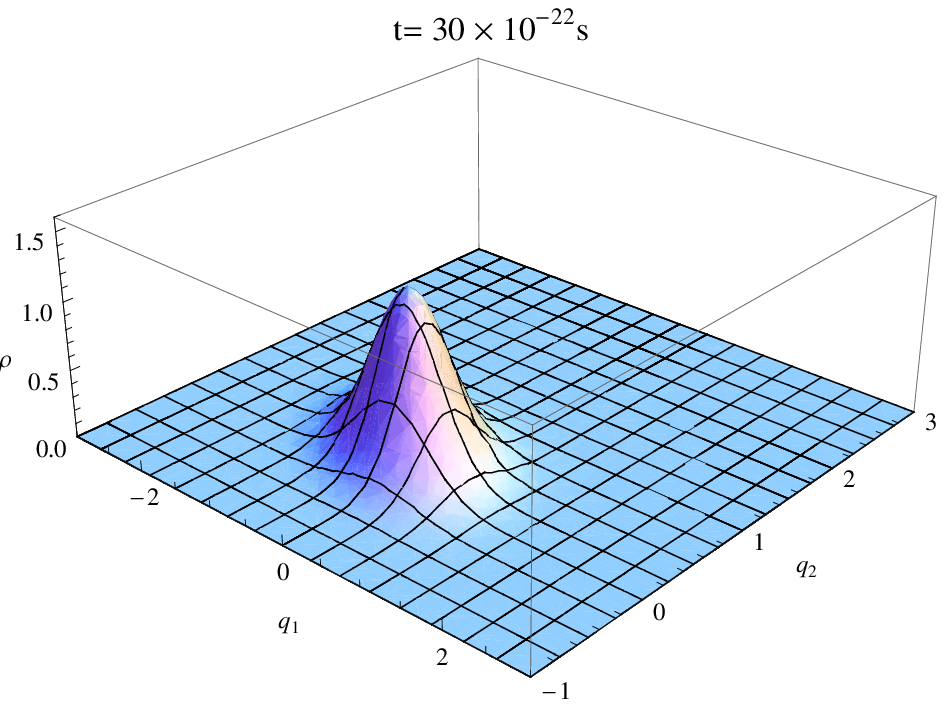}}
\resizebox*{0.40\textwidth}{!}{\includegraphics{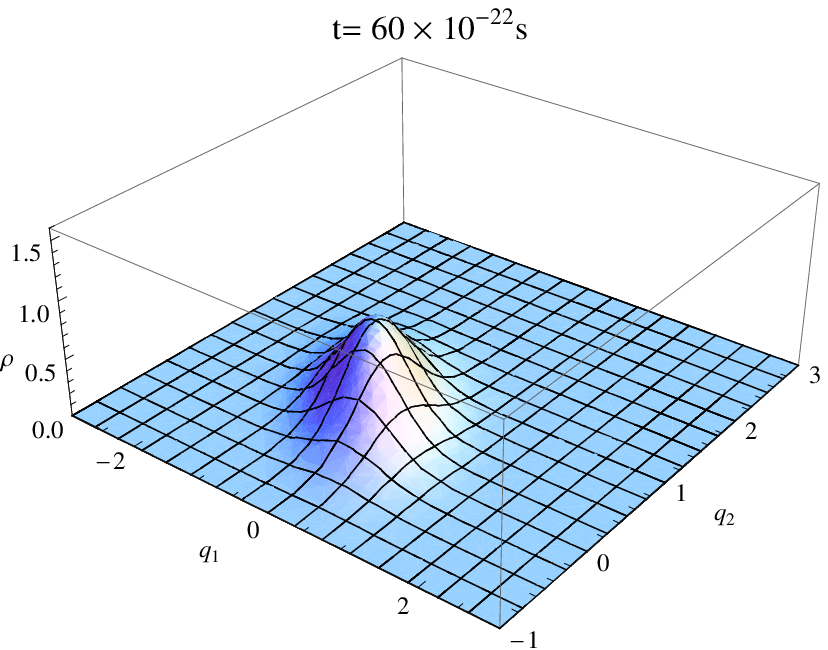}}
\par}}

\caption{\label{fig9} (Color online) The probability density function $\rho(q_1,q_2;t)$ 
at different values of time. The parameters are $m_{1}= 2.5 \hbar^2/{\rm MeV}$, $m_{2}= 60 \hbar^2/{\rm MeV}$, $\hbar\omega_1=1.7$ {\rm MeV}, $\hbar\omega_2=0.6$ {\rm MeV}, $\nu_{12}=7$ {\rm  MeV}, $\hbar \lambda_{11}=2.5$ {\rm MeV}, $\hbar \lambda_{22}=0.6$ {\rm MeV}, $T=0.1$ {\rm MeV}, $\sigma_{q_1q_1}(0)=0.4$, $\sigma_{p_1p_1}(0)=\hbar^2/(4\sigma_{q_1q_1}(0))$, $\sigma_{q_2q_2}(0)=7\times 10^{-2}$, $\sigma_{p_2p_2}(0)=\hbar^2/(4\sigma_{q_2q_2}(0))$, $\sigma_{q_1}(0)=-6$, $\sigma_{q_2}(0)=0$, $\sigma_{p_1}(0)=9\hbar$, $\sigma_{p_2}(0)=0$; all other parameters are set to zero. These  values correspond to an initial total energy  $E_0=-115.9\ {\rm MeV}$.}
\end{figure}
 The solutions of the equations of motion for this case may be obtained by simply making the replacement
\begin{equation}
  \omega_{k}\rightarrow i\omega_k
  \end{equation}
  in the expression of the matrix $M$ [see equation~(\ref{matr})]. In the subsequent  calculations we use the diffusion coefficients~(\ref{co1})-(\ref{co6}). In general, the  density matrix in coordinates space is given by the integral of the Wigner distribution function of the system with respect to momentum variables~\cite{dis}, namely,
 \begin{equation}
  \rho(q_1,q_2,\cdots q_N;t)=\int_{-\infty}^{\infty} \int_{-\infty}^{\infty}\cdots \int_{-\infty}^{\infty} \prod_{k=1}^N dp_k  W(q_1,p_1,q_2,p_2\cdots q_N,p_N;t).\label{pro}
  \end{equation}
  It has been shown using path integral techniques that for the harmonic oscillator, if the initial density matrix is Gaussian, then it remains Gaussian at any moment of the time~\cite{adam1}. This means that the Wigner function can be determined by a simple substitution of  the time-dependent variances and expectation values, namely,
  \begin{equation} W(q_1,p_1,q_2,p_2\cdots q_N,p_N;t)=(2 \pi)^{-\tfrac{N}{2}}\sqrt{\det(\sigma^{-1}(t))}\exp\Bigl[-\frac{1}{2}(Z-{\mathcal V(t)})^T\sigma^{-1}(t)(Z-{\mathcal V}(t))\Bigl],
\end{equation}
with $Z=\{q_1,p_1,q_2,p_2\cdots q_N,p_N\}^T$, and $\sigma^{-1}(t)$ denotes the inverse  of $\sigma(t)$. In figure~\ref{fig9} we display the time development of the probability density function~(\ref{pro}) for the case of a two-dimensional parabolic barrier. One can see that at short times $\rho(q_1,q_2,t)$ gets sharper as compared with its initial shape;  obviously, this is accompanied by an increase of its height since its area should be constant. (more precisely, the integral of the above quantity over the whole space should be equal to unity.) The height of the distribution function  decreases with time whereas $\rho(q_1,q_2,t)$ spreads in coordinates space, to become centered around $(\sigma_{q_1}(\infty),\sigma_{q_2}(\infty))$ at sufficiently long times. 

The probability of finding  the packet to the right of the barrier in the $q_1$ direction is given by (note that the top of the barrier is located at the origin)
\begin{equation}
P(t)=\int\limits_{-\infty}^{\infty}dq_2\int\limits_0^{\infty}dq_1 \rho(q_1,q_2;t)=\int\limits_{-\infty}^{\infty}dp_2\int\limits_{-\infty}^{\infty}dp_1\int\limits_{-\infty}^{\infty}dq_2\int\limits_0^{\infty}dq_1 W(q_1,p_1,q_2,p_2;t).
\end{equation}
  $P(t)$ is used here to  quantify the penetrability through the parabolic barrier. Notice that in one dimension, enhancement of the tunneling was found for large values of the friction coefficient~\cite{adam1}. Here we shall investigate the effect of dissipation when other degrees of freedom are considered. 
  
\begin{figure}[htba]
{\centering{
\resizebox*{0.50\textwidth}{!}{\includegraphics{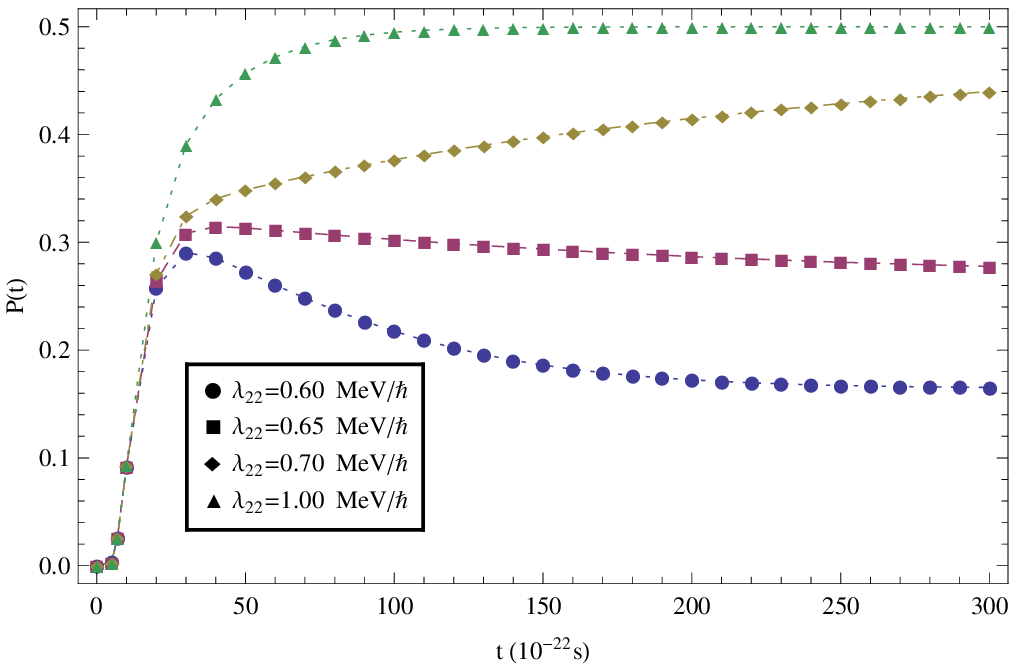}}
\par}}

\caption{\label{fig10} (Color online)  Probability of penetration through the barrier $P(t)$ as a function of time for different values of the friction coefficient $\lambda_{22}$. The parameters are $m_{1}= 2.5 \hbar^2/{\rm MeV}$, $m_{2}= 60 \hbar^2/{\rm MeV}$, $\hbar\omega_1=1.7$ {\rm MeV}, $\hbar\omega_2=0.6$ {\rm MeV}, $\nu_{12}=7$ {\rm  MeV}, $\hbar \lambda_{11}=2.5$ {\rm MeV}, $T=0.1$ {\rm MeV}, $\sigma_{q_1q_1}(0)=0.4$, $\sigma_{p_1p_1}(0)=\hbar^2/(4\sigma_{q_1q_1}(0))$, $\sigma_{q_2q_2}(0)=7\times 10^{-2}$, $\sigma_{p_2p_2}(0)=\hbar^2/(4\sigma_{q_2q_2}(0))$, $\sigma_{q_1}(0)=-6$, $\sigma_{q_2}(0)=0$, $\sigma_{p_1}(0)=9\hbar$, $\sigma_{p_2}(0)=0$; all other parameters are set to zero. These  values correspond to an initial total energy  $E_0=-115.9\ {\rm MeV}$.}
\end{figure}

  In figure~\ref{fig10}, the  penetration probability is shown as a function of time for different values of the friction coefficient $\lambda_{22}$. It can bee seen that in the interval $0\leq t\leq 20\times 10^{-22} \ {\rm s}$, $P(t)$ is almost the same for all values of $\lambda_{22}$. Then the curves spread apart from each other, to tend to certain  asymptotic values which  depend, in turn, on the dissipation rate. Indeed, we see that the greater the value of the friction coefficient, the larger the asymptotic penetrability, as clearly indicated in figure~\ref{fig10}. Thus  the tunneling through the barrier  in the $q_1$ direction is enhanced by the dissipation in the other degree of freedom. Quite surprisingly, we find that for sufficiently large values of $\lambda_{22}$,  the wave packet is trapped near the top of the barrier ($P(t)\to0.5$).
  
  We have also studied  the dependence of the tunneling on the coupling constant $\nu_{12}$. It turns out that $P(t)$ is inversely proportional to the latter parameter. We also found that the penetrability increases with the temperature which can be explained by the increase of the values of the diffusion coefficients.
\section{Summary}
In this paper we have axiomatically derived the multidimensional diffusion coefficients for a set of $N$ coupled harmonic oscillators using Lindblad's approach. The only assumption we have made is the existence of a Gibbs steady state for the system under consideration. It turns out that the general form of the off-diagonal elements of the diffusion matrix is given by the arithmetic mean of two terms, each having the form of one of the diffusion coefficients corresponding to the one-dimensional harmonic oscillator. Furthermore, we have shown that the fluctuation-dissipation theorem holds for both diagonal and off-diagonal coefficients in momentum coordinates. The elements of the friction tensor are found to be not independent. This is due to the fundamental constraints on the diffusion coefficients. We have derived the equations of motion for the expectation values and variances, and solved them for arbitrary values of the coupling strengths, without having recourse to any perturbative treatment.  This is, indeed, one of the advantages of the investigated model. We have applied our results to the description of mass and charge asymmetry coordinates in deep-inelastic collisions. We find that the expectation values of the coordinates and momenta  do not depend on the temperature, in contrast to the variances which are temperature dependent. The decay of these quantities is faster for large values of both the friction coefficients, and the coupling  constants. The correlation between the degrees of freedom is more appreciable when the coupling is strong. This was confirmed by comparing the theoretical  results with the experimental data. It is also shown that dissipation in one degree of freedom enhance the tunneling in sub-barrier processes. In conclusion, the model is quite interesting, in the sense that it is exactly solvable;  further extensions and investigations may be carried out.       

 \begin{appendix}
 \section{}
 The result of applying  the superoperator $\mathcal{L}$ [see equation~(\ref{super})] to the operator $\hat V_\ell^\dag\hat V_\ell$ is given by
 \begin{eqnarray}
 {\mathcal L}[\hat V_\ell^\dag\hat V_\ell]&=&\sum_k\Biggl\{\Bigl[|a_k^{\ell}|^2 \cosh^2(\hbar\beta\omega_k)-\frac{i}{m_k\omega_k} (a^{\ell*}_kb_k^\ell+a^\ell_k b_k^{\ell*})\cosh(\hbar\beta\omega_k)\sinh(\hbar\beta\omega_k)\nonumber\\&-&\frac{|b^\ell_k|^2}{m_k^2\omega_k^2}\sinh^2(\hbar\beta\omega_k)\Bigr]\hat p_k^2+\Bigl[a^{\ell*}_k b^\ell_k\cosh^2(\hbar\beta\omega_k)+i\Bigl(|a^{\ell}_k|^2 m_k \omega_k-\frac{|b_k^\ell|^2}{m_k\omega_k}\Bigr)\nonumber\\&\times&\cosh(\hbar\beta\omega_k)\sinh(\hbar\beta\omega_k)+a^\ell_k b^{\ell*}_k\sinh^2(\hbar\beta\omega_k)\Bigl]\hat p_k\hat q_k+
 \Bigl[a^{\ell}_k b^{\ell*}_k\cosh^2(\hbar\beta\omega_k)\nonumber\\
 &+&i\Bigl(|a^{\ell}_k|^2 m_k \omega_k-\frac{|b_k^\ell|^2}{m_k\omega_k}\Bigr)\cosh(\hbar\beta\omega_k)\sinh(\hbar\beta\omega_k)+a^{\ell*}_k b^{\ell}_k\sinh^2(\hbar\beta\omega_k)\Bigl]\hat p_k\hat q_k\nonumber\\&+& \Bigl[|b_k^{\ell}|^2 \cosh^2(\hbar\beta\omega_k)+i m_k\omega_k (a^{\ell}_kb_k^{\ell*}+a^{\ell*}_k b_k^{\ell})\cosh(\hbar\beta\omega_k)\sinh(\hbar\beta\omega_k)\nonumber\\&-&|a^\ell_k|^2m_k^2\omega_k^2\sinh^2(\hbar\beta\omega_k)\Bigr]\hat q_k^2\Biggr\}+\sum_{k\neq j}\Biggl\{\Bigl[a^{\ell*}_k a^\ell_j\cosh(\hbar\beta\omega_k)\cosh(\hbar\beta\omega_j)\nonumber\\&-&i\frac{a^{\ell*}_k b^\ell_j}{m_j\omega_j}\cosh(\hbar\beta\omega_k)\sinh(\hbar\beta\omega_j)-i\frac{b^{\ell*}_ka^\ell_j}{m_k\omega_k}\cosh(\hbar\beta\omega_j)\sinh(\hbar\beta\omega_k)\nonumber\\&-&\frac{b^{\ell*}_k b^\ell_j}{m_km_j\omega_k\omega_j}\sinh(\hbar\beta\omega_k)\sinh(\hbar\beta\omega_j)\Bigl]\hat p_k\hat p_j
 +\Bigl[b^{\ell*}_k b^\ell_j\cosh(\hbar\beta\omega_k)\cosh(\hbar\beta\omega_j)\nonumber\\&+&i b^{\ell}_k a^\ell_j m_j\omega_j\cosh(\hbar\beta\omega_k)\sinh(\hbar\beta\omega_j)+i a^{\ell*}_k b^\ell_j m_k\omega_k\cosh(\hbar\beta\omega_j)\sinh(\hbar\beta\omega_k)\nonumber\\&-&a^{\ell*}_k a^\ell_jm_k m_j\omega_k\omega_j\sinh(\hbar\beta\omega_k)\sinh(\hbar\beta\omega_j)\Bigl]\hat q_k\hat q_j+\Bigl[(a^{\ell*}_kb^\ell_j+a^\ell_k b^\ell_j)\cosh(\hbar\beta\omega_j)\nonumber\\ &\times&\cosh(\hbar\beta\omega_k)+i(a^{\ell*}_k a^\ell_j+a^\ell_ka^{\ell*}_j)m_j\omega_j\cosh(\hbar\beta\omega_k)\sinh(\hbar\beta\omega_j)-\frac{i}{m_k\omega_k}\nonumber\\&\times& (b^{\ell*}_k b^\ell_j+b^\ell_k b^{\ell*}_j)\sinh(\hbar\beta\omega_k)\cosh(\hbar\beta\omega_j)+\frac{i m_j\omega_j}{m_k\omega_k}(b^{\ell*}_k a^\ell_j+b^\ell_k a^{\ell*}_j)\sinh(\hbar\beta\omega_k)\nonumber\\&\times&\sinh(\hbar\beta\omega_j)\Bigl]\hat p_k\hat q_j\Biggr\}
 \end{eqnarray}
 \end{appendix}

\end{document}